\documentclass[12pt]{article}
\usepackage{graphicx,amsmath,amssymb,cite}

\newlength{\dinwidth}
\newlength{\dinmargin}
\renewcommand{\vec}[1]{\boldsymbol{#1}}
\newcommand{\slashed}[1]{\makebox[0pt][l]{/}#1}
\newcommand{\dif}{\mathrm{d}}
\newcommand{\diff}[1]{\frac{\mathrm{d}#1}{#1}}
\newcommand{\xB}{x_{\scriptscriptstyle{B}}}
\newcommand{\Pom}{{\hspace{ -0.1em}I\hspace{-0.2em}P}}

\newcommand{\xPom}{x_\Pom}
\newcommand{\ad}{a^{\rm D}}
\newcommand{\ap}{a^\Pom}
\newcommand{\fd}{F_2^{{\rm D}(3)}}

\begin{document}
\titlepage
\begin{flushright}
  IPPP/05/07      \\
  DCPT/05/14      \\
  DESY 05-055     \\
  21st July 2005  \\
\end{flushright}

\vspace*{0.5cm}

\begin{center}
  
  {\Large \bf Diffractive parton distributions from perturbative QCD}

  \vspace*{1cm}

  \textsc{A.D. Martin$^a$, M.G. Ryskin$^{a,b}$ and G. Watt$^c$} \\

  \vspace*{0.5cm}

  $^a$ Institute for Particle Physics Phenomenology, University of Durham, DH1 3LE, UK \\
  $^b$ Petersburg Nuclear Physics Institute, Gatchina, St.~Petersburg, 188300, Russia \\
  $^c$ Deutsches Elektronen-Synchrotron DESY, 22607 Hamburg, Germany

\end{center}

\vspace*{0.5cm}

\begin{abstract}
  The asymptotic collinear factorisation theorem, which holds for diffractive deep-inelastic scattering, has important modifications in the sub-asymptotic HERA regime.  We use perturbative QCD to quantify these modifications.  The diffractive parton distributions are shown to satisfy an inhomogeneous evolution equation.  We emphasise that it is necessary to include both the gluonic and sea-quark $t$-channel components of the perturbative Pomeron.  The corresponding Pomeron-to-parton splitting functions are derived in the Appendix.
\end{abstract}

\section{Introduction} \label{sec:introduction}

\begin{figure}
  \centering
    \includegraphics[width=0.4\textwidth]{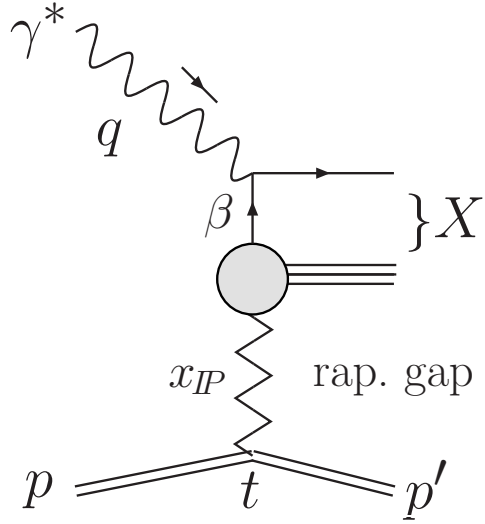}
  \caption{Diagram showing the kinematic variables which describe the DDIS process $\gamma^* p\to X + p$: $\xPom$ is the fraction of the proton's momentum transferred through the rapidity gap, $\beta\equiv \xB/\xPom$ is the fraction of this momentum carried by the struck quark, $\xB$ is the Bjorken $x$ variable, $q^2 \equiv -Q^2$ is the photon virtuality, and $t$ is the squared 4-momentum transfer.}
  \label{fig:variables}
\end{figure}

A notable feature of deep-inelastic scattering is the existence of diffractive events, $\gamma^* p\to X + p$, in which the slightly deflected proton and the cluster $X$ of outgoing hadrons are well-separated in rapidity.\footnote{The process and the corresponding variables are shown in Fig.~\ref{fig:variables}.}  At high energies, the large rapidity gap is believed to be associated with Pomeron, or vacuum quantum number, exchange.  Some secondary Reggeons also have vacuum quantum numbers, but these contributions are exponentially suppressed as a function of the gap size, and are negligible at small $\xPom$.  The diffractive events make up an appreciable fraction of all (inclusive) deep-inelastic events, $\gamma^* p \to X$.  We will refer to the diffractive and inclusive processes as DDIS and DIS respectively.

The recent improvement in the precision of the DDIS data \cite{ZEUS,H1data} allow improved analyses to be performed and more reliable diffractive parton densities to be extracted. Moreover, measurements of diffractive charm production in DIS are available \cite{Chekanov:2003gt}, which provide added constraints, particularly on the diffractive gluon density.

It is common to perform analyses of DDIS data based on two levels of factorisation.  First the diffractive structure function $F_2^{{\rm D}(3)}$ may be written as the convolution of the usual coefficient functions as in DIS with diffractive parton distribution functions (DPDFs) $\ad$ \cite{Collins:1997sr}:
\begin{equation} \label{eq:DDISfac}
  F_2^{{\rm D}(3)} = \sum_{a=q,g} C_{2,a}\otimes\ad,
\end{equation}
with factorisation scale $\mu_F$, where $\ad=\beta q^{\rm D}$ or $\beta g^{\rm D}$ satisfy DGLAP evolution in $\mu_F$.  The collinear factorisation theorem \eqref{eq:DDISfac} applies when $\mu_F$ is made large, therefore it is correct up to power-suppressed corrections.  In a second stage, Regge factorisation is usually assumed \cite{Ingelman:1984ns} (see, for example, the preliminary H1 analysis \cite{H1data}), such that the diffractive parton densities $\ad$ are written as a product of the Pomeron flux factor $f_\Pom(\xPom,t)$ and the Pomeron parton densities $a^\Pom=\beta q^\Pom$ or $\beta g^\Pom$.  Taking $\mu_F=Q$, the $t$-integrated form is
\begin{equation} \label{eq:resolvedPom}
  \ad(\xPom,\beta,Q^2) = f_\Pom(\xPom)\,a^\Pom(\beta,Q^2),
\end{equation}
where the Pomeron flux factor is taken from Regge phenomenology,
\begin{equation} \label{eq:H1Pomflux}
  f_\Pom(\xPom) = \int_{t_\mathrm{cut}}^{t_\mathrm{min}}\!\dif{t}\frac{\mathrm{e}^{B_\Pom\,t}}{\xPom^{2\alpha_\Pom(t)-1}},
\end{equation}
with $\alpha_\Pom(t) = \alpha_\Pom(0) + \alpha_\Pom^\prime\,t$.  For simplicity of presentation we omit the contribution of secondary Reggeons to the right-hand side of \eqref{eq:resolvedPom}. Strictly speaking, the parameters $\alpha_\Pom(0)$, $\alpha_\Pom^\prime$, and $B_\Pom$ must be taken from fits to soft hadron data.  However, then one fails to describe the $\xPom$ dependence of the $F_2^{{\rm D}(3)}$ diffractive data.  Therefore, as a rule (see \cite{H1data}), $\alpha_\Pom^{\prime}$ and $B_\Pom$ are fixed by the analyses of soft hadron data, while $\alpha_\Pom(0)$, and the parameters describing the input Pomeron parton distributions at some scale $\mu_0$, are determined from a fit to the DDIS data.  In these analyses, the Pomeron is treated as an effective pole in the complex angular momentum plane, and regarded as a hadron-like object of more or less fixed size.  This Regge factorisation takes place in the non-perturbative region at some low scale $\mu$, with $\mu<\mu_0$, of the order of the (inverse) size of the hadron. However, in such fits, the value of $\alpha_\Pom(0)$ extracted from DDIS data (for example, 1.17 in Ref.~\cite{H1data}) lies significantly above the value of 1.08 obtained from soft hadron data \cite{Donnachie:1992ny}.  This indicates that there is a contribution coming from the small-size component of the Pomeron, that is, coming from the perturbative QCD region where the vacuum singularity has a larger intercept.

This Regge factorisation approach is a simplified phenomenological model.  Here we shall not assume Regge factorisation for the whole $\ad$, but instead study the impact of applying perturbative QCD to the analysis of DDIS data.  The procedure we discuss in this paper formed the basis of our earlier DDIS analyses \cite{MRW1,MRW2}.

The content of this paper is as follows. In Section \ref{sec:collfact} we recall the collinear factorisation which underlies the description of inclusive DIS.  This paves the way for the discussion in Section \ref{sec:evdpdf} of the evolution equations for the diffractive parton distributions $\ad$ due to the perturbative Pomeron.  These equations contain an appreciable inhomogeneous term\footnote{The existence of the inhomogeneous term has been known for some time \cite{Ryskin:1990fb,Levin:1992bz}, but it is ignored in most phenomenological analyses of DDIS data.}, analogous to the inhomogeneous term in the evolution equations for the parton distributions of the photon.  Since the evolution equations for the diffractive densities $\ad$ are a little subtle, it may be helpful at this stage to look ahead to the discussion in Section \ref{sec:discussion}.  Although collinear factorisation holds in DDIS in the asymptotic limit, at the relevant HERA energies there exist important modifications.  These are discussed in Section \ref{sec:factddis}.  Then in Section \ref{sec:seapom} we explain why it is necessary to include a Pomeron made of a sea-quark--antiquark pair in addition to the Pomeron made of two $t$-channel gluons.  Section \ref{sec:pqcdanal} shows some relevant properties of a recent analysis \cite{MRW1} using this approach, and Section \ref{sec:discussion} summarises how universal diffractive parton densities are extracted, after allowing for the modifications to DDIS factorisation.  The Appendix contains a derivation of all of the Pomeron-to-parton leading order (LO) splitting functions that are necessary to analyse DDIS data.

\section{Collinear factorisation} \label{sec:collfact}

The problem in the analysis of both DIS and DDIS data is that we can only use perturbative QCD (pQCD) at small distances, that is, at large $Q^2$. Within pQCD we can study the evolution of parton distributions, but the initial distributions at some relatively low scale $\mu$ are of non-perturbative origin and, at present, have to be determined by fitting to the data.  A factorisation theorem underlies the analysis.  It enables the amplitude to be factored into two parts, one purely in the pQCD domain, and the other parameterised by a phenomenological ansatz. In terms of Feynman diagrams, the factorisation is based on the resummation of the series of the most important higher-order corrections where the small coupling $\alpha_S$ is enhanced by a large $\ln(Q^2/\mu^2)$.  That is, we can divide such diagrams, at a `logarithmic loop or cell', into a part depending only on large scales from a part containing the low scale.  Let us be more explicit.

\begin{figure}
  \centering
    \includegraphics[width=0.4\textwidth]{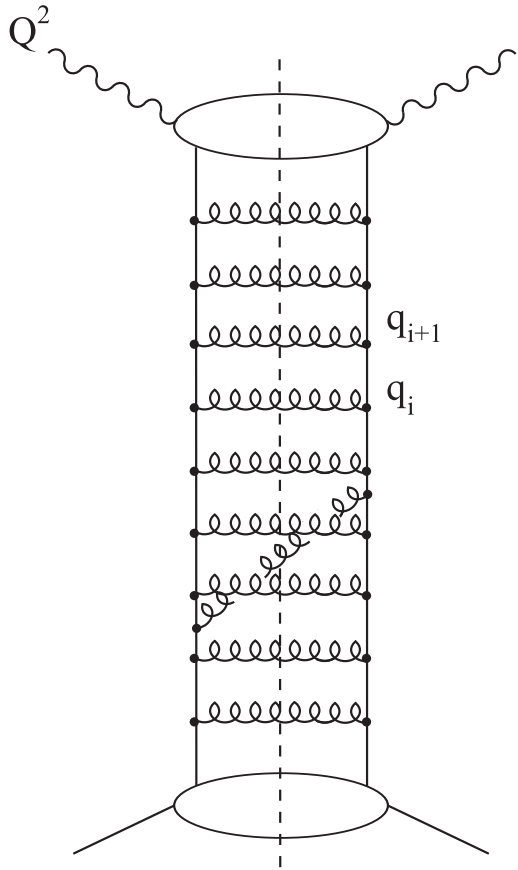}
  \caption{Ladder-type diagram for the DIS cross section. The gluon line crossing two gluon rungs means that this is a NNLO contribution with two $\alpha_S$ factors unaccompanied by $\ln(Q^2/\mu^2)$ enhancements.}
  \label{fig:ladder}
\end{figure}

First, recall how collinear factorisation occurs in DIS. In a physical gauge (such as the axial gauge for the gluon field with $\mathcal{A}^A_\mu {q^\prime}^\mu=0$, where $q^\prime$ is the light-cone vector in the photon direction) the leading log contributions come from ladder-type graphs\footnote{Besides the ladder graphs, we have to include virtual loop corrections, which may be included in the usual way by the plus prescription.}; see Fig.~\ref{fig:ladder} \cite{Dokshitzer:1977sg}.  In such graphs any box may be considered as the logarithmic loop which provides the factorisation at some scale $\mu = \mu_F$.  Indeed, the integral over each virtuality $q_i$ takes the form $\int^{q^2_{i+1}} \dif{q^2_i}/q^2_i$, and in order to generate a large log to compensate the small $\alpha_S$, we need to be in the strongly-ordered region\footnote{(LO) DGLAP evolution effectively sums up these leading $\ln(Q^2/\mu^2)$ terms. In the case of BFKL we have an analogous factorisation.  However, instead of $\int \dif{q_i^2}/{q_i^2}$, we have $\int \dif{z_i}/z_i$ to provide the strong ordering in emission angles of the gluons.  In this case the $t$-channel partons are two reggeised gluons.} with $q^2_i \ll q^2_{i+1}$.  Because of this strong ordering, the parton $q_i$ may be considered on-mass-shell for the upper part of the diagram, which can be regarded as the matrix element of the hard subprocess where all the other virtualities are much larger than $q_i$.  This upper part represents the $\gamma^* q_i$ interaction, and contains the coefficient function and the upper part of the ladder.  At LO we keep the LO coefficient function and the LO splitting functions, so each upper loop is logarithmic and could provide an alternative factorisation scale $\mu_F$.  The distribution of parton $q_i$ depends on the parameterised input distribution at some fixed scale $\mu_0$ determined by fitting to the data in the pQCD domain. Starting with the phenomenological distribution at $\mu_0$ the PDF is evolved up to the factorisation region $\mu_F$.  The physical result does not depend on the choice of the factorisation scale $\mu_F$ (up to $\mathcal{O}(\alpha_S)$ corrections when working at LO accuracy). If we change $\mu_F$ we move from one cell to another.  The change in the distribution in going to the new $\mu_F$ is compensated by the change of the hard interaction in the upper part of the diagram.

Corrections to this LO picture occur if two gluons may have comparable virtualities or if an extra gluon were to cross one or more of the existing gluon rungs (as shown in Fig.~\ref{fig:ladder}).  In these cases the large log integrations are absent.  In such contributions the extra powers of $\alpha_S$ are unaccompanied by logarithmic enhancement, that is, they are higher-order in $\alpha_S$ (NLO, NNLO, \ldots) contributions to the splitting functions.  If the additional gluon couples to the upper blob or the uppermost rung, then it would involve higher-order contributions to the coefficient function.

\section{Evolution of the diffractive parton densities} \label{sec:evdpdf}

Let us now turn to DDIS. Here we have to include a rapidity gap between the ladder and the proton target, and to identify the appropriate factorisation scale for the DPDF evolution.  

\subsection{Starting scales}

\begin{figure}
  \centering
 \includegraphics[width=1.0\textwidth]{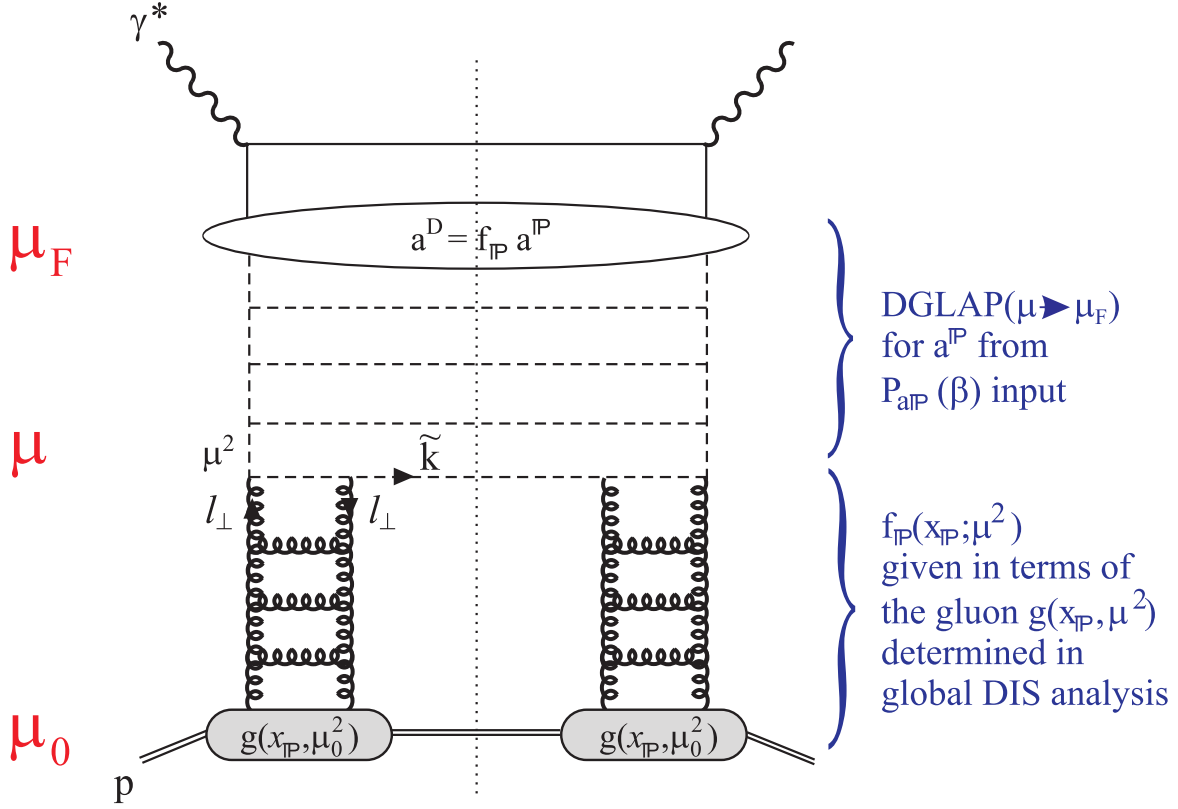}
  \caption{A ladder-type diagram showing a contribution to the diffractive parton densities $\ad(\xPom,\beta,\mu_F^2)$ in the perturbative region, $\mu > \mu_0\sim 1$ GeV.  DGLAP evolution for the Pomeron densities ($\ap = \beta q^\Pom,\beta g^\Pom$) is performed from $\mu$ to $\mu_F$ for each component $\mu$ of the perturbative Pomeron in the perturbative interval $\mu_0<\mu <\mu_F$ and then the sum is taken.  The virtuality $\mu^2$ of the first $t$-channel parton in the upper ladder is fixed by the momentum $\tilde{k}$ of the emitted parton at the edge of the rapidity gap, $\mu^2=k_t^2/\tilde{\beta}$; see \eqref{eq:ladderVirtuality}.  The dashed lines in the upper ladder may be either gluons or quarks.  The two gluon ladders shown in the diagram represent Pomerons.  In general, these ladders can also contain quarks as well as gluons.  We label the contributions where the two uppermost $t$-channel partons (labelled $l_\perp$ on the left-hand side) are gluons, in both lower ladders, by $\Pom=G$.  There are also contributions in which both ladders have a sea-quark--antiquark pair as the two uppermost $t$-channel partons.  We denote these contributions by $\Pom=S$.  The flux factor in this case, $f_{\Pom=S}$, is given by \eqref{eq:source} with $g$ replaced by the sea-quark density of the proton, $S$.  There are also interference terms in which one ladder is $\Pom=G$ and the other ladder is $\Pom=S$.  We denote this contribution by $\Pom=GS$.  The factorisation scale $\mu_F^2$ is usually taken to be $Q^2$.
}
  \label{fig:DDISladder}
\end{figure}

Consider the LO Feynman graph where Pomeron exchange is described by a gluon ladder; see the left-hand side of Fig.~\ref{fig:DDISladder}.  The kinematics of the process are fixed by the momentum $\tilde{k}$ of the first emitted parton at the end of the rapidity gap.  Indeed, the virtuality of the first $t$-channel parton in the upper ladder fixes the scale $\mu^2=k_t^2/\tilde{\beta}$, where $\tilde{\beta}$ is the (light-cone) momentum fraction of the Pomeron carried by parton $\tilde{k}$ and $k_t$ is its transverse momentum.  The derivation of $\mu^2=k_t^2/\tilde{\beta}$ is given in the Appendix; see \eqref{eq:ladderVirtuality}.  We assume that $|t| \ll k_t^2$.  This scale $\mu$ plays the r\^ole of the lowest factorisation scale for the usual DGLAP evolution in the upper part of the diagram.  \emph{Simultaneously}, it is the upper scale for the lower part of the diagram.  Indeed, the integral over the transverse momentum $l_t$ of the $t$-channel gluon has the logarithmic form for $l_t^2 \ll \mu^2$, whereas for $l_t^2 \gg \mu^2$ it converges as $\dif{l_t^2}/l_t^4$ and its contribution may be neglected as usual.  After accounting for the evolution in the lower part of the diagram, and integrating over $l_t$, the probability amplitude to find the appropriate $t$-channel gluons is given, at LO, by the conventional\footnote{To be precise, the \emph{skewed} gluon distribution is required, which can be written as a constant factor \eqref{eq:Rg} multiplied by the conventional gluon distribution \cite{Shuvaev:1999ce}.} integrated gluon distribution of the proton, $x_\Pom g(x_\Pom,\mu^2)$, which is known from global analyses of DIS and related hard-scattering data.

The evolution of $\ad$ is a little subtle.  Before we integrate over the momentum $k_t$ --- that is, selecting events with a fixed transverse momentum of the lowest parton --- we can indeed see that the scale $\mu$ is the lowest possible factorisation scale for the diffractive parton densities $\ad$.  However, for inclusive DDIS, we must integrate over $k_t$.  The integral over $k_t$ translates into an integral over $\mu$, and may be written, up to a normalisation factor, in the form\footnote{The derivation of the form of \eqref{eq:pertflux} is given in Section \ref{sec:two-gluon-exchange} of the Appendix.}
\begin{equation}
\ad = \int_{\mu_0^2}^{Q^2}\diff{\mu^2} \frac{1}{\xPom}\left[\frac{\alpha_S}{\mu} \xPom g(\xPom,\mu^2)\right]^2 \ap.
\label{eq:pertflux}
\end{equation}
The term $[...]^2/\xPom$ plays the r\^ole of the Pomeron flux, $f_\Pom$, which occurs in the conventional analyses; see \eqref{eq:resolvedPom} and \eqref{eq:H1Pomflux}. At first sight, the integral is concentrated in the infrared region of low $\mu$.  However, for DDIS we consider very small $x_\Pom$.  In this domain the gluon has a large anomalous dimension.  Asymptotically, BFKL predicts $x_\Pom g(x_\Pom,\mu^2) \sim (\mu^2)^{0.5}$ for fixed $\alpha_S$ \cite{Lipatov:1996ts}. In this case the integral \eqref{eq:pertflux} takes a logarithmic form, and we cannot neglect the large scale $\mu$ contribution.  This complicates the evolution by adding a non-negligible {\it inhomogeneous} term to the DGLAP equation for the diffractive parton densities $\ad$.

\subsection{Inhomogeneous contribution to the evolution equation for $\ad$}

Before we present the evolution equation for DDIS, it is informative to recall the evolution equation for the parton distributions of the photon, which also contains an inhomogeneous term.  It is of the form \cite{Witten:1977ju}
\begin{equation}
\frac{\partial q(x,Q^2)}{\partial\ln Q^2}=\frac{\alpha_S}{2\pi}\int_x^1\diff{z} P_{qq}(z)\,q\left(\frac{x}{z},Q^2\right)~+~\frac{\alpha_{\rm em} e_q^2}{2\pi} \int_x^1\diff{z} P_{q\gamma}(z)\,\gamma\left(\frac{x}{z},Q^2\right),
\label{eq:photonDGLAP}
\end{equation}
where, for simplicity, we show only the quark non-singlet evolution.  The right-hand side of the evolution equation now contains an inhomogeneous term, which acts as an {\it extra source} of quarks produced by the splitting of a point-like photon.  The point-like photon has a distribution of the form
\begin{equation}
\gamma(y,Q^2)=\delta(1-y).
\end{equation}
Note that, since the photon is point-like, the extra source is independent of $Q^2$.

We also have an \emph{extra source} of contributions in the evolution of the diffractive parton densities, $\ad$, for $\mu$ in the perturbative region. Thus the evolution equation for $\ad$ contains an inhomogeneous term.  This term depends on the scale $\mu$.  From \eqref{eq:pertflux}, we see that the dependence is described by the factor
\begin{equation}
  f_\Pom(\xPom;\mu^2)\equiv \frac{N}{\xPom}\left[\frac{\alpha_S}{\mu} x_\Pom g(x_\Pom,\mu^2)\right]^2,
\label{eq:source}
\end{equation}
which specifies the strength of the {\it inhomogeneous source} of partons coming from the component of the Pomeron of size $1/\mu$.  (Our choice of the normalisation factor $N$ is specified in the Appendix; see \eqref{eq:f2d3GdipoleTPomflux}.)  To be more precise, we mean partons coming from the component of the Pomeron wave function corresponding to the integration of the lower parts of the diagram shown in Fig.~\ref{fig:DDISladder} over $l_t$ up to scale $\mu$.  An explicit pQCD calculation \cite{Wusthoff:1997fz} shows that the corresponding splitting functions of this perturbative Pomeron into quarks and gluons are of the form
\begin{align}
P_{q\Pom}(\beta)&~\propto~\beta^3 (1-\beta), \label{eq:spfn}\\
P_{g\Pom}(\beta)&~\propto~(1+2\beta)^2 (1-\beta)^2.
\label{eq:spfn2}
\end{align}
The derivation and normalisation\footnote{Of course, only the product of the flux factor and the splitting functions have a precisely defined normalisation.  Our choice for the normalisation of the separate factors is given in the Appendix.} of the splitting functions (and flux factor $f_\Pom$) are given in the Appendix; see \eqref{eq:f2d3GdipoleTSigmaPom} and \eqref{eq:gG}.  Differentiating \eqref{eq:pertflux} with respect to $\ln Q^2$, we see that the evolution equations for the diffractive parton densities ($\ad=\beta q^{\rm D}$, $\beta g^{\rm D}$) are
\begin{align}
\frac{\partial \ad}{\partial \ln Q^2}
&=\int_{\mu_0^2}^{Q^2}\diff{\mu^2}\;f_\Pom(\xPom;\mu^2)\frac{\partial a^\Pom}{\partial\ln Q^2}~+~\left.f_\Pom(\xPom;\mu^2)\,a^\Pom(\beta,Q^2;\mu^2)\right|_{\mu^2=Q^2} \notag\\
&=\int_{\mu_0^2}^{Q^2}\diff{\mu^2}\;f_\Pom(\xPom;\mu^2)\;\frac{\alpha_S}{2\pi}\sum_{a^\prime=q,g}P_{aa^\prime}\otimes{a^\prime}^{\Pom}~+~f_\Pom(\xPom;Q^2)\,a^\Pom(\beta,Q^2;Q^2)\notag\\
&=\frac{\alpha_S}{2\pi}\sum_{a^\prime=q,g}P_{aa^\prime}\otimes{a^\prime}^{\rm D}~+~f_\Pom(\xPom;Q^2)\,P_{a\Pom}(\beta).
\label{eq:DDISDGLAP}
\end{align}
Here, $a^\Pom(\beta,Q^2;\mu^2)$ are the Pomeron parton densities DGLAP-evolved from a starting scale $\mu^2$ up to $Q^2$, from input distributions $a^\Pom(\beta,\mu^2;\mu^2)=P_{a\Pom}(\beta)$.

If we were to assume that $f_\Pom$ were independent of $\mu^2$, as is the case for BFKL asymptotics, then \eqref{eq:DDISDGLAP} would be an inhomogeneous DGLAP equation exactly analogous to that for the photon.  The only essential difference would be the form of the $\beta$ dependence of the splitting functions $P_{a\Pom}$.  At first sight this appears strange.  The Pomeron, unlike the photon, is an extended object, and we might have anticipated a form factor dependence $f_\Pom \sim 1/(\mu^2 R^2_\Pom)$. For such an extreme behaviour, the inhomogeneous term would be a power correction and could perhaps be neglected in the evolution equation.  However, as $\xPom\to 0$,  the proton gluon density $g(\xPom,\mu^2)$ grows as $(\mu^2)^{0.5}$, and compensates the form factor power-like suppression.  The HERA domain is an intermediate region: the anomalous dimension $\gamma$ is not small, but is less than 0.5, with $g\sim (\mu^2)^{\gamma}$.  Thus, although the integral in \eqref{eq:pertflux} is convergent at large $\mu^2$, we still cannot neglect the inhomogeneous term in \eqref{eq:DDISDGLAP}.

\begin{figure}
  \centering
  \includegraphics[width=0.8\textwidth,clip]{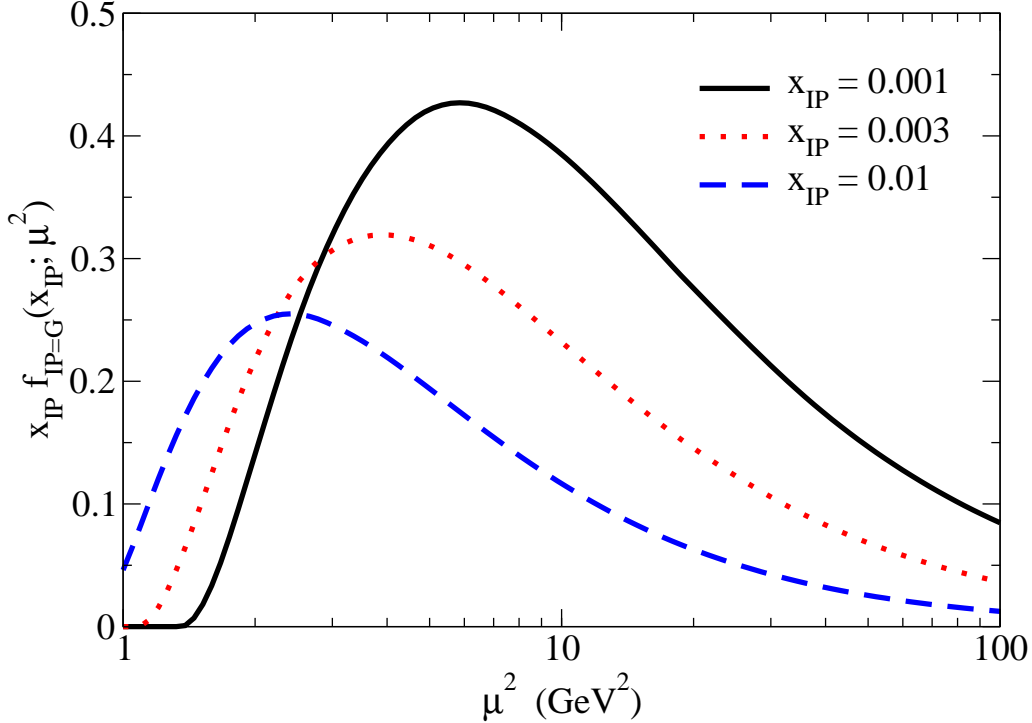}
  \caption{The $\mu^2$ dependence of the flux factor $f_\Pom$, given by \eqref{eq:source} with $N=1$, for three different values of $\xPom $ and using the MRST2001 NLO gluon distribution of the proton \cite{MRST2001}.}
  \label{fig:f}
\end{figure}

The behaviour of the inhomogeneous term can be seen from Fig.~\ref{fig:f}, which shows the $\mu^2$ dependence of the flux $f_\Pom$ of \eqref{eq:source} with $N=1$ for three typical HERA values of $\xPom$. Of course, $f_\Pom$ is not constant and dies out with increasing $\mu^2$, but it is still appreciable up to rather large $\mu^2$.  Therefore, we cannot neglect the inhomogeneous contribution when we describe HERA diffractive data with $Q^2 \sim 10$--$50$ GeV$^2$.  Typically the perturbative contribution, from scales $\mu>\mu_0\sim 1$ GeV, is found \cite{MRW1} to be responsible for about half or more of the DDIS HERA data, depending on the value of $\beta$.  Moreover, it is seen from Fig.~\ref{fig:f} that, even for very large $Q^2$, DDIS factorisation must be modified when working with input scales $\mu_0^2$ less than about 10--50 ${\rm GeV}^2$.   Also, note that the convergence of the $\mu^2$ integral decreases with decreasing $\xPom$. 

In practice, rather than to solve the inhomogeneous equation \eqref{eq:DDISDGLAP} directly, it is more convenient to use a separate standard homogeneous DGLAP equation for each component $\mu$ of the Pomeron.  Then we can sum up (that is, integrate over $\mu$) the different $\mu^2$ contributions, in which the standard DGLAP evolution of each component starts from its own scale $\mu$, provided that $\mu$ is in the pQCD domain ($\mu > \mu_0$), and continues up to the (collinear) factorisation scale $\mu_F$.  The contribution coming from $\mu<\mu_0$ must be treated non-perturbatively.  This component of the diffractive densities $\ad$ is included in the starting distributions at $\mu_0$ whose parameters are obtained by fitting to data.

\section{Factorisation in DDIS and sub-asymptotic effects} \label{sec:factddis}

From the formal viewpoint, for any fixed $\xPom$, the inhomogeneous term in the evolution equation \eqref{eq:DDISDGLAP} for $\ad$ dies out as $Q^2 \to \infty$.  Thus if we evolve from a large enough starting scale, $\mu_S$, we are entirely in the perturbative regime and the diffractive parton densities, $\ad$, are described by the usual (homogeneous) DGLAP equations and satisfy the collinear factorisation theorem, just like the parton densities for DIS whose factorisation was described in Section \ref{sec:collfact}.  The factorisation theorem for DDIS was proved in Ref.~\cite{Collins:1997sr}.

However, this is not true in the HERA domain, and in this section we emphasise those places where a more precise treatment of diffractive parton densities is necessary.  Nevertheless, universal diffractive parton distributions can still be obtained from analysing DDIS data, provided that care is taken in the analysis.

\subsection{Inclusion of the inhomogeneous term and the effect on $g^{\rm D}$}

For a low starting scale $\mu_S$, but still satisfying $\mu_S \gg \Lambda_{\rm QCD}$, we can no longer omit the inhomogeneous term in the evolution \eqref{eq:DDISDGLAP} of the diffractive parton densities, $\ad$.  One consequence is that we will obtain a smaller diffractive gluon density  $g^{\rm D}$.  Indeed, it is known from the global analyses of DIS data that the gluon is mainly driven by $\partial F_2/\partial\ln Q^2 \sim \alpha_S\,g$.  However, for DDIS, the perturbative Pomeron contribution to $\fd$ is \cite{MRW1}
\begin{align}
  F_{2,{\rm pert.}}^{{\rm D}(3)}(\xPom,\beta,Q^2) &= \int_{\mu_0^2}^{Q^2}\diff{\mu^2}\;f_{\Pom}(\xPom;\mu^2)\;F_{2}^\Pom(\beta,Q^2;\mu^2)\label{eq:f2d3pert}\\
  \Longrightarrow\quad\frac{\partial F_{2,{\rm pert.}}^{{\rm D}(3)}}{\partial \ln Q^2} &= \int_{\mu_0^2}^{Q^2}\diff{\mu^2}\;f_{\Pom}(\xPom;\mu^2)\frac{\partial F_{2}^\Pom(\beta,Q^2;\mu^2)}{\partial\ln Q^2} + f_{\Pom}(\xPom;Q^2)F_{2}^\Pom(\beta,Q^2;Q^2). \label{eq:f2d3pertdiff}
\end{align}
Here, $F_{2}^\Pom(\beta,Q^2;\mu^2)$ is the Pomeron structure function, evaluated from the DGLAP-evolved Pomeron parton densities $a^\Pom(\beta,Q^2;\mu^2)$.  The first term of \eqref{eq:f2d3pertdiff} behaves roughly as $\alpha_S\,g^{\rm D}$.  Therefore, part of the derivative $\partial \fd/\partial \ln Q^2$ comes from the upper limit $Q^2$ of the integral over the Pomeron scale $\mu$, and so this results in a smaller diffractive gluon density than if the second term of \eqref{eq:f2d3pertdiff} was neglected.

\subsection{Heavy quark contributions} \label{sec:heavy}

To be specific we consider the contributions of the charm quark, $c$.  For moderate $Q^2$ it is convenient to use the fixed flavour number scheme, where the charm contribution arises from photon--gluon fusion $\gamma^* g^\Pom \to c\bar{c}$.  However, in this case we will miss the diagram where the Pomeron directly produces the $c$ quark, that is, when $c$ is the lowest parton $\tilde{k}$ in the upper ladder of Fig.~\ref{fig:DDISladder}. In this case there is no evolution and no factorisation.  This contribution should be added separately.

At high $Q^2$ the difference between light and heavy quarks disappears.  Then we should use a variable flavour number scheme, where a diffractive charm density is introduced at scales above the charm threshold.

Here is a good place to emphasise the difference between the Pomeron densities $\ap$ and the $t$-channel content of the Pomeron.  Already in the charm example above, we see that although the Pomeron is built up only of gluons, it produces not only gluons, but also light and heavy quarks. 

In analogy to heavy quark production, the {\it direct} coupling of the perturbative Pomeron has to be included in the description of high $E_T$ dijets in DDIS.  Again, this cannot be described purely as a convolution of $\ad$ with a `hard' matrix element.  This \emph{direct} contribution should be added separately.

\subsection{Twist-four contributions in DDIS}

From \eqref{eq:spfn} and \eqref{eq:spfn2}, we see that the leading-twist splitting functions vanish as $\beta \to 1$.  For large $\beta$ and the $Q^2$ values typical at HERA, we cannot neglect the twist-four $F_L^{{\rm D}(3)}$ contribution, which goes to a constant value as $\beta \to 1$.   This contribution was calculated in Ref.~\cite{Wusthoff:1997fz}, and turns out to be numerically appreciable.  Moreover, it dies out with increasing $Q^2$ rather slowly since the extra $1/Q^2$ twist-four suppression is partly compensated by the sizeable anomalous dimension $\gamma$ of the gluon; recall $[g(\xPom,Q^2)]^2 \sim (Q^2)^{2\gamma}$.  This twist-four contribution goes beyond the leading twist-two approximation.  It is described by its own twist-four evolution and coefficient functions.

\section{The sea-quark component of the Pomeron} \label{sec:seapom}

\begin{figure}
  \centering
    \includegraphics[width=0.8\textwidth,clip]{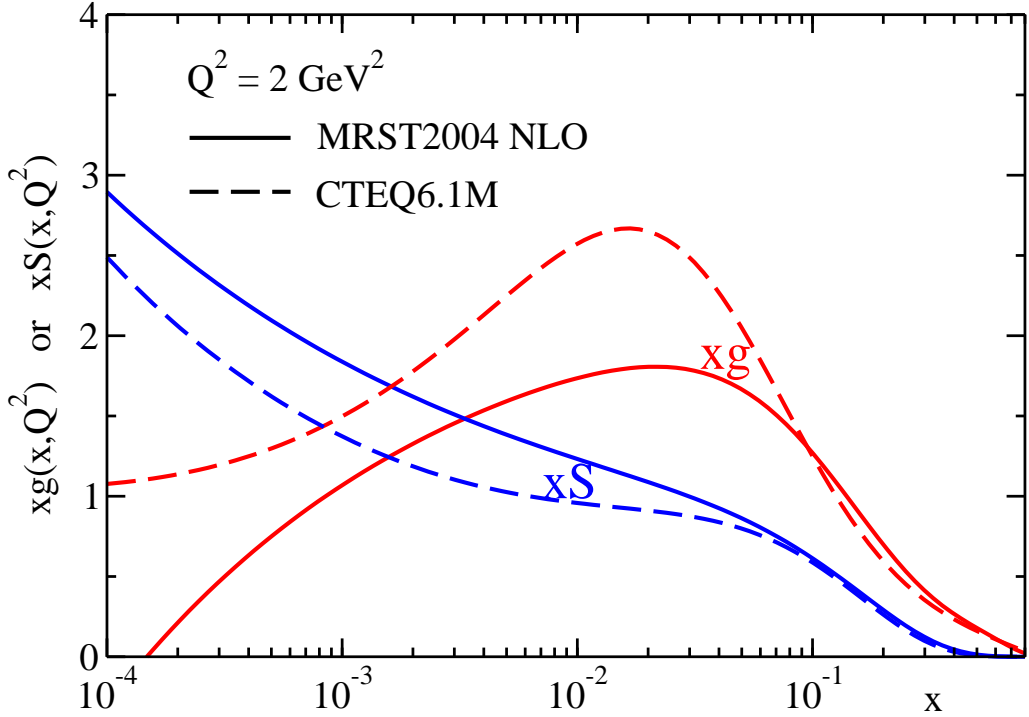}
  \caption{The behaviour of the gluon and sea-quark distributions at $Q^2=2$ GeV$^2$ found in the CTEQ6.1M \cite{Pumplin:2002vw} and MRST2004 NLO \cite{Martin:2004ir} global analyses.  The valence-like behaviour of the gluon is evident.}
  \label{fig:gS}
\end{figure}

So far we have taken the Pomeron to be a parton ladder where the two uppermost $t$-channel partons are gluons.  However, at small scales $\mu^2$, the gluon densities have a valence-like structure.  They decrease with decreasing $x$ already from $x \sim 0.01$ for $\mu^2 \sim 2$ GeV$^2$; see Fig.~\ref{fig:gS}.  On the other hand, the sea-quark density $S=2(\bar{u}+\bar{d}+\bar{s})$ increases as $xS \sim x^{-0.2}$ with decreasing $x$.  As a consequence we have to include another contribution to the Pomeron in which the two uppermost $t$-channel partons in the lower ladders in Fig.~\ref{fig:DDISladder} are a sea-quark--antiquark pair.  Thus we must introduce a sea-quark Pomeron flux $f_{\Pom=S}$ given by \eqref{eq:source} with $\xPom g$ replaced by $\xPom S$.  To avoid confusion we denote the flux in \eqref{eq:source} by $f_{\Pom=G}$.  In general, the two lower ladders in Fig.~\ref{fig:DDISladder}, shown as gluon ladders, contain both quarks and gluons.  For a gluonic Pomeron ($\Pom=G$) or a sea-quark Pomeron ($\Pom=S$) the two uppermost partons in the ladders are gluons or sea quarks respectively.  In addition, we must include the interference contribution coming from one $\Pom=G$ ladder and one $\Pom=S$ ladder in Fig.\ref{fig:DDISladder}.  We denote this contribution by $\Pom=GS$.

We present the calculation of all the LO Pomeron-to-parton splitting functions in the Appendix.  The calculation of the splitting functions for the gluonic Pomeron (see $P_{q,\Pom=G}$ and $P_{g,\Pom=G}$ of \eqref{eq:spfn} and \eqref{eq:spfn2} respectively) already exist\footnote{Here we clarify certain factors of 2.} in the literature \cite{Wusthoff:1997fz}, but those for the sea-quark Pomeron are derived for the first time.\footnote{These forms were used in the analyses of Ref.~\cite{MRW1,MRW2}.}  They have the forms
\begin{align}
P_{q,\Pom=S}(\beta)&~\propto~\beta (1-\beta), \label{eq:qSa}\\
P_{g,\Pom=S}(\beta)&~\propto~(1-\beta)^2;
\label{eq:gSa}
\end{align}
see \eqref{eq:qS} and \eqref{eq:gS} of the Appendix.  In addition, there are interference contributions between the gluonic Pomeron and the sea-quark Pomeron, which we label by the notation $\Pom=GS$.  These splitting functions have the forms
\begin{align}
P_{q,\Pom=GS}(\beta)&~\propto~\beta^2 (1-\beta), \label{eq:qGSa}\\
P_{g,\Pom=GS}(\beta)&~\propto~(1-\beta)^2(1+2\beta);
\label{eq:gGSa}
\end{align}
see \eqref{eq:qGS} and \eqref{eq:gGS} of the Appendix.

\section{pQCD analysis of DDIS data} \label{sec:pqcdanal}

\begin{figure}
  \centering
  \includegraphics[width=\textwidth,clip]{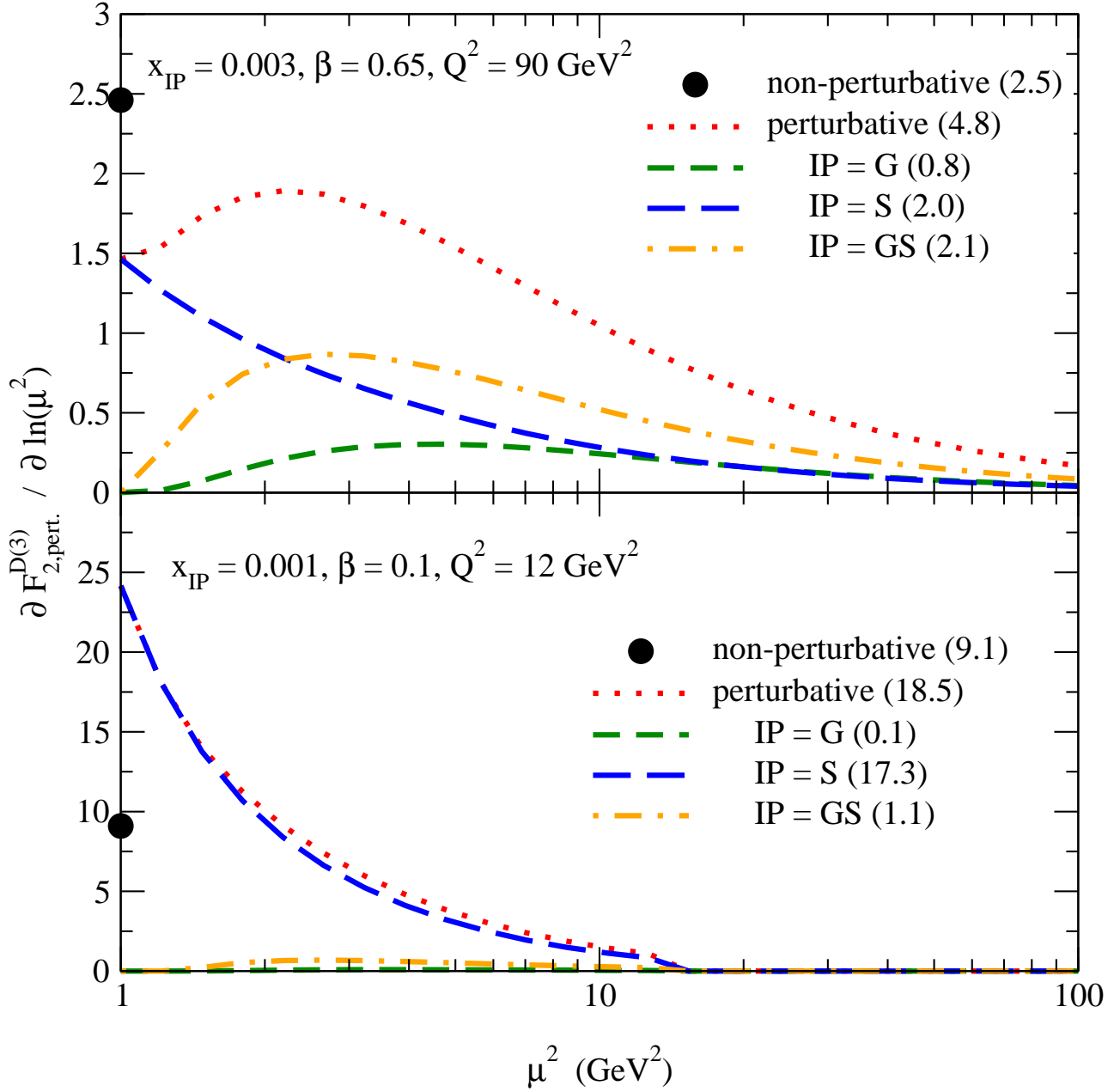}
  \caption{The perturbative ($\Pom=G$,$S$, and the $GS$ interference) and non-perturbative contributions to $\fd$, for two sets of $\xPom$, $\beta$, and $Q^2$ values, found in the analysis of HERA DDIS data in Ref.~\cite{MRW1}.  The plots show the $\mu^2$ dependence of the perturbative contributions; their integral over $\mu^2$ is shown by the numbers in parentheses in the legend.}
  \label{fig:pnp}
\end{figure}

The pQCD approach described above has been used to analyse the new HERA DDIS data \cite{MRW1}.  A good description of both the ZEUS and H1 data was obtained.  The plots in Fig.~\ref{fig:pnp} show the separate non-perturbative and perturbative ($\Pom=G$,$S$, and the $GS$ interference) contributions to $\fd$, for two sets of values of $\xPom$, $\beta$, and $Q^2$.  The numbers in brackets in Fig.~\ref{fig:pnp} show the total contribution of the different components starting from $\mu_0=1$ GeV; and correspond in the perturbative cases to taking the integral over $\mu^2$.  We see, first, that the perturbative contribution is significant and sometimes dominant and, second, that the modifications of DDIS factorisation are important, that is, the inhomogeneous term in the evolution equation plays a crucial r\^ole.

\begin{figure}
  \centering
    \includegraphics[width=\textwidth]{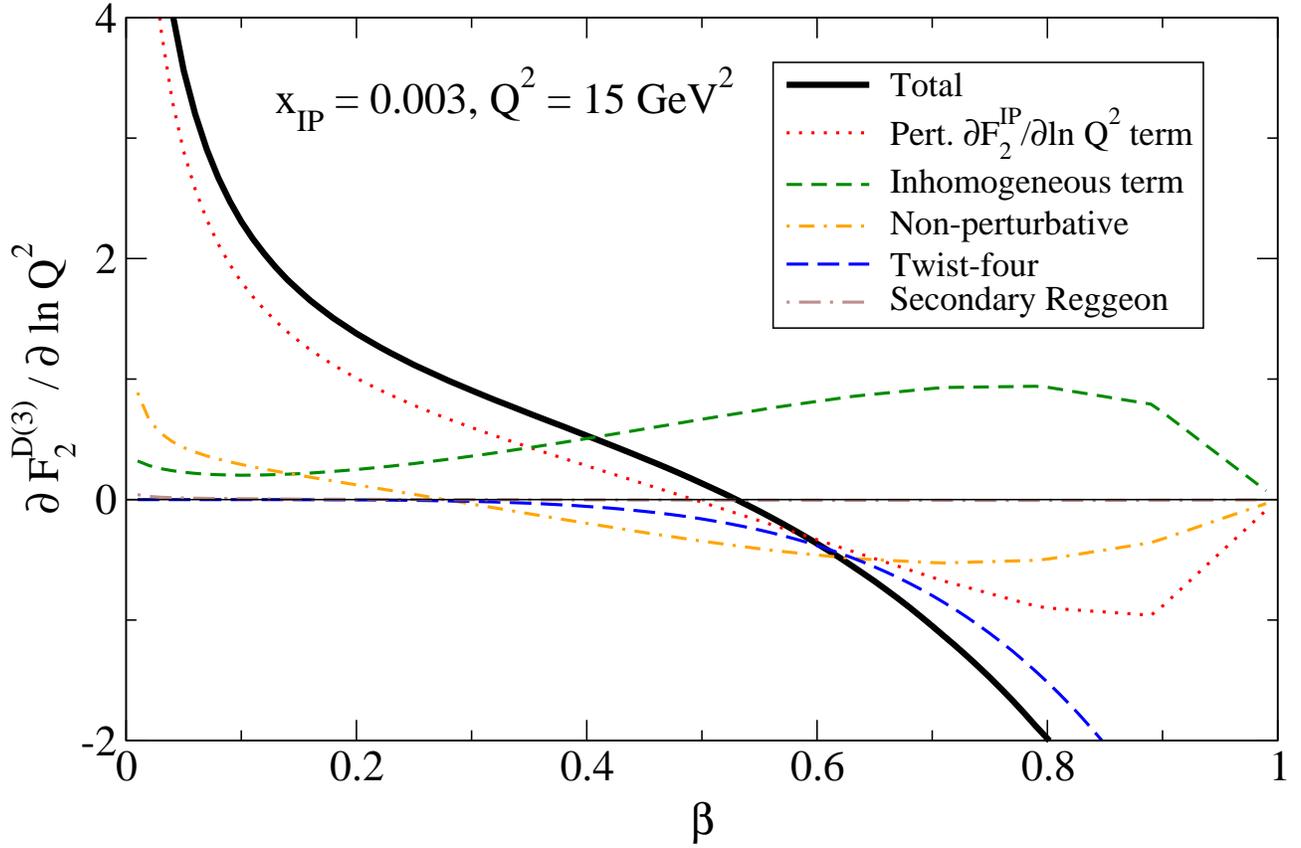} 
  \caption{The breakdown of the contributions to the description of the slope $\partial \fd/\partial\ln Q^2$ as a function of $\beta$, for $Q^2=15$ GeV$^2$ and $\xPom=0.003$ obtained in the MRW analysis \cite{MRW1} of the combined ZEUS and H1 DDIS data, which uses MRST2001 NLO partons \cite{MRST2001} to calculate the perturbative Pomeron flux.}
  \label{fig:slope}
\end{figure}

It is informative to study the various contributions to $\partial \fd/\partial\ln Q^2$ in some detail, since this is the quantity which mainly drives the behaviour of the diffractive gluon distribution.  Recall that in the global analyses of inclusive DIS data, the small-$x$ gluons are extracted using DGLAP evolution in which $\partial F_2/\partial\ln Q^2 \sim \alpha_S\,g$.  As we have mentioned, in DDIS, the derivative $\partial \fd/\partial\ln Q^2 $ contains an additional contribution arising from the inhomogeneous term; see \eqref{eq:f2d3pertdiff}.  In Fig.~\ref{fig:slope} we show the $\beta$ dependence of the $\ln Q^2$ derivative of $\fd$, for $\xPom=0.003$ and $Q^2=15$ GeV$^2$, obtained in the recent MRW analysis \cite{MRW1} which takes $\mu_0=1$ GeV.  From this figure, we see that the major contribution to $\partial \fd/\partial\ln Q^2$ comes from the first term in \eqref{eq:f2d3pertdiff}, that is, the perturbative DGLAP contribution for $\mu>\mu_0$ given by the dotted curve.  The non-perturbative contribution coming from $\mu<\mu_0$ is rather small for $\beta \lesssim 0.5$.  The contribution corresponding to the differentiation of the upper limit of the $\mu^2$ integral, that is, the second term in \eqref{eq:f2d3pertdiff}, is shown by the short dashed line.  Clearly it is not negligible. The twist-four ($F_L^{{\rm D}(3)}$) contribution is important, as expected, for large $\beta$, and the contribution from the secondary Reggeon at this small value of $\xPom$ is practically invisible.  In summary, we see that the slope for $\beta \lesssim 0.2$ is dominated by the perturbative DGLAP contribution, and for $\beta \gtrsim 0.8$ by the twist-four $F_L^{{\rm D}(3)}$ contribution, while for $\beta \sim 0.5$ the inhomogeneous contribution is largest.  In any complete analysis of DDIS data, it is clear that the inhomogeneous term must be taken into account.

\section{Discussion} \label{sec:discussion}

The QCD analysis of diffractive structure functions is subtle.  To clarify the situation it may be helpful to pursue the analogy with the parton distributions of the photon.  The evolution can be expressed in two alternative forms.  In the first form, the DGLAP evolution \eqref{eq:photonDGLAP} of the parton distributions of the photon includes an inhomogeneous term arising from the direct splitting of a point-like photon into a $q \bar{q}$ pair.  This splitting may take place at any current scale.  An alternative way to account for this effect is to treat the photon as a parton, and to consider homogeneous DGLAP equations which embrace the evolution of the gluon and quark distributions, together with that of the photon itself.  For such a photon density we have the trivial distribution $\gamma=\delta(1-x)$, but now the {\it inhomogeneous} term is embodied in the enlarged set of {\it homogeneous} DGLAP equations for $q$, $g$ together with $\gamma$.  The advantage of the latter approach is that, in such a form, we have enlarged the subspace where the factorisation theorem applies.  We can now factor off the matrix element of the hard subprocess $i+j \to X$ where $i,j=q$, $g$, or $\gamma$.  Thus we have the possibility of the photon directly participating in the hard interaction.

In the case of the diffractive parton distributions, we have a very similar situation to that of the parton densities of the photon.  The DGLAP evolution contains an inhomogeneous contribution arising from the Pomeron-to-parton splitting, as given by the last term in \eqref{eq:DDISDGLAP}.  In the $Q^2 \to \infty$ limit, and at fixed $\xPom$, due to the $1/Q^2$ factor explicit in $f_\Pom$, this may be treated as a power correction and safely neglected.  Then the DDIS factorisation theorem is valid.  However, for small $\xPom$, the parton densities increase rapidly with increasing $Q^2$, partly neutralising the power suppression of the inhomogeneous Pomeron-induced term.  As a consequence we must retain the inhomogeneous term, and we are forced to reject the attractive postulate that DDIS factorisation\cite{Collins:1997sr} holds down to low scales.\footnote{This postulate is the basis of the description of DDIS that is presented in Ref.~\cite{Brodsky:2004hi}.  We thank Paul Hoyer for clarifying discussions concerning this work.}

Since, in practice, it is easier to work with homogeneous DGLAP equations, we can proceed in one of two ways.  First, just as in the photon case, we may treat the Pomeron as an extra parton, and enlarge the set of DGLAP equations to include $f_\Pom$, as well as $q^{\rm D}$ and $g^{\rm D}$.  We denote the Pomeron density as $f_\Pom$ to be consistent with our previous notation.  The extra terms $P_{q\Pom}f_\Pom$ and $P_{g\Pom}f_\Pom$ in the homogeneous DGLAP equations for $q^{\rm D}$ and $g^{\rm D}$ involve the Pomeron-to-parton splitting functions.  Moreover, just as in the photon case, to calculate the cross section we have to include the direct Pomeron coupling to the hard subprocess (see Section \ref{sec:heavy} for an example).  In this approach the effective Pomeron distribution $f_\Pom$ is now described by the Pomeron flux $f_\Pom \equiv f_{\Pom=G}$ of \eqref{eq:source}, together with the corresponding flux $f_{\Pom=S}$ for the sea-quark component of the Pomeron, and the $\Pom=GS$ interference term.  Recall that the inclusive parton densities of the proton, $\xPom g$ and $\xPom S$, which occur in these fluxes, satisfy their own DGLAP evolution equations.

Equally well, the inhomogeneous DGLAP evolution of diffractive parton densities could be accomplished in an alternative way.  We may sum up the parton densities given by a series of homogeneous DGLAP equations with different values of the starting scale $\mu$.  Suppressing the $\xPom$ and $\beta$ dependence, the general structure is
\begin{equation}
\ad(\mu_F;\mu_0)=\ad_{\rm non-pert.} (\mu_F;\mu_0)~+~\int^{\mu_F^2}_{\mu_0^2}\diff{\mu^2}\;f_\Pom(\mu^2)\;\ap(\mu_F;\mu),
\label{eq:homo}
\end{equation}
where $\ap(\mu_F;\mu)$ results from DGLAP evolution up to $\mu_F$ starting from the input $\ap(\mu;\mu)=P_{a\Pom}(\beta)$.  The non-perturbative term, $\ad_{\rm non-pert.}$, is exactly analogous to the procedure used in the Regge factorisation analysis described in Section \ref{sec:introduction}, but with $\alpha_\Pom(0)$ fixed at the `soft' value of 1.08 \cite{Donnachie:1992ny}.  The important new ingredient is the pQCD contribution given by the second term on the right-hand side of \eqref{eq:homo}, which has contributions from the gluonic and sea-quark Pomeron ($\Pom=G$,$S$), and their interference ($\Pom=GS$).  Setting $\mu_F=Q$, it is clear that \eqref{eq:homo} satisfies the inhomogeneous evolution equation \eqref{eq:DDISDGLAP}.

In summary, we have shown how to obtain \emph{universal} diffractive parton densities $\ad$ which can be used in the description of different diffractive processes.  Of course, for diffractive production in high energy hadron--hadron collisions, we have to take care that the rapidity gaps are not populated by secondaries produced during the soft interaction of spectators.  This well-known rapidity gap survival factor, usually denoted by $S^2$ \cite{Bjorken}, may be calculated from phenomenological models tuned to describe elastic and related `soft' hadron--hadron processes \cite{Khoze:2000wk,Gotsman:1999xq}.  

In addition to the hard subprocesses originating from the collision of quarks and gluons from the \emph{resolved} Pomeron, we must also include the contribution where the perturbative Pomeron {\it directly} participates in the hard interaction.  In particular, the calculation in the Appendix of the cross section for diffractive open charm production may be treated as the coefficient function for $\gamma^*\Pom \to c\bar{c}$ in the fixed flavour number scheme.

\section*{Acknowledgements}

We thank Markus Diehl for drawing our attention to the need to give a detailed description of the procedure used to analyse DDIS, and for valuable comments on the text.  ADM thanks the Leverhulme Trust for an Emeritus Fellowship.  This work was supported by the UK Particle Physics and Astronomy Research Council, by the Federal Program of the Russian Ministry of Industry, Science and Technology (grant SS-1124.2003.2), by the Russian Fund for Fundamental Research (grant 04-02-16073), and by a Royal Society Joint Project Grant with the former Soviet Union.

\appendix
\setcounter{equation}{0}
\renewcommand{\theequation}{A.\arabic{equation}}

\section{Appendix: The $gg$ and $q\bar{q}$ Pomeron in pQCD}

In this Appendix we compute the lowest-order Feynman diagrams which give the $\beta$ dependence of the quark singlet and gluon starting distributions of the perturbative Pomeron, $\Sigma^\Pom(\beta,\mu^2;\mu^2)$ and $g^\Pom(\beta,\mu^2;\mu^2)$.  Here, the quark singlet distribution $\Sigma^{\Pom} = u^\Pom + d^\Pom + s^\Pom + \bar{u}^\Pom + \bar{d}^\Pom + \bar{s}^\Pom$, with all six quark densities of the Pomeron equal to each other.  In Section \ref{sec:two-gluon-exchange} we compute $\Sigma^{\Pom=G}$ and $g^{\Pom=G}$ assuming that the QCD Pomeron is made from two $t$-channel gluons.  Wherever possible, we check our results with those of W\"usthoff \cite{Wusthoff:1997fz} in the limit of massless quarks and with Ref.~\cite{Levin:1996vf} in the case of massive quarks; see also the original work of Ref.~\cite{Nikolaev:1990ja}.  Then in Section \ref{sec:two-quark-exchange} we extend the formalism to calculate $\Sigma^{\Pom=S}$ and $g^{\Pom=S}$ assuming that the Pomeron is represented by a $t$-channel sea-quark--antiquark pair.  Finally, in Section \ref{sec:interference} we calculate $\Sigma^{\Pom=GS}$ and $g^{\Pom=GS}$ resulting from the interference between the gluonic Pomeron and the sea-quark Pomeron.  As explained in the main text, the starting distributions, $\Sigma^\Pom(\beta,\mu^2;\mu^2)$ and $g^\Pom(\beta,\mu^2;\mu^2)$, can be thought of as Pomeron-to-parton splitting functions, $P_{q\Pom}$ and $P_{g\Pom}$ respectively.

We will work in the leading logarithmic approximation (LLA) to derive the factorised form of \eqref{eq:f2d3pert}:
\begin{equation}
  F_{2,{\rm pert.}}^{{\rm D}(3)}(\xPom,\beta,Q^2) = \int_{\mu_0^2}^{Q^2}\diff{\mu^2}\;f_{\Pom}(\xPom;\mu^2)\;F_{2}^\Pom(\beta,Q^2;\mu^2).\label{eq:appf2d3pert}
\end{equation}
Higher order corrections (for example, to NLO accuracy) may be incorporated by considering proton and Pomeron parton densities satisfying NLO DGLAP evolution, by using the NLO coefficient functions to calculate $F_2^\Pom$, and by computing $\mathcal{O}(\alpha_S)$ corrections to the Pomeron-to-parton splitting functions.  However, note that the factorised structure of \eqref{eq:appf2d3pert} persists even at NLO (and beyond).

\subsection{Two-gluon exchange ($\Pom=G$)} \label{sec:two-gluon-exchange}

\begin{figure}
  \centering
  \begin{minipage}{0.49\textwidth}
    \hspace{0.5\textwidth}(a)\\
    \includegraphics[width=\textwidth]{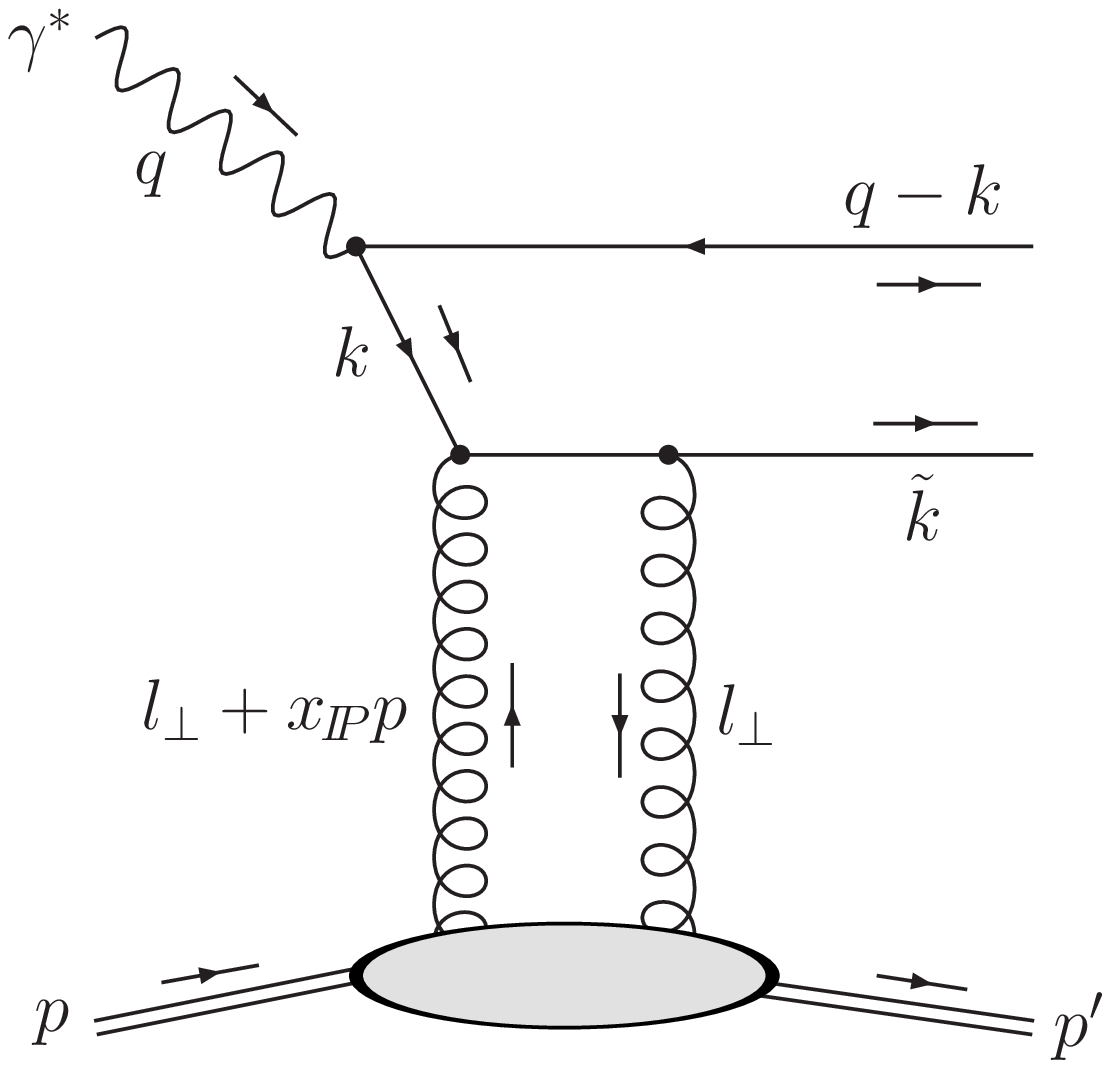}
  \end{minipage}
  \begin{minipage}{0.49\textwidth}
    \hspace{0.5\textwidth}(b)\\
    \includegraphics[width=\textwidth]{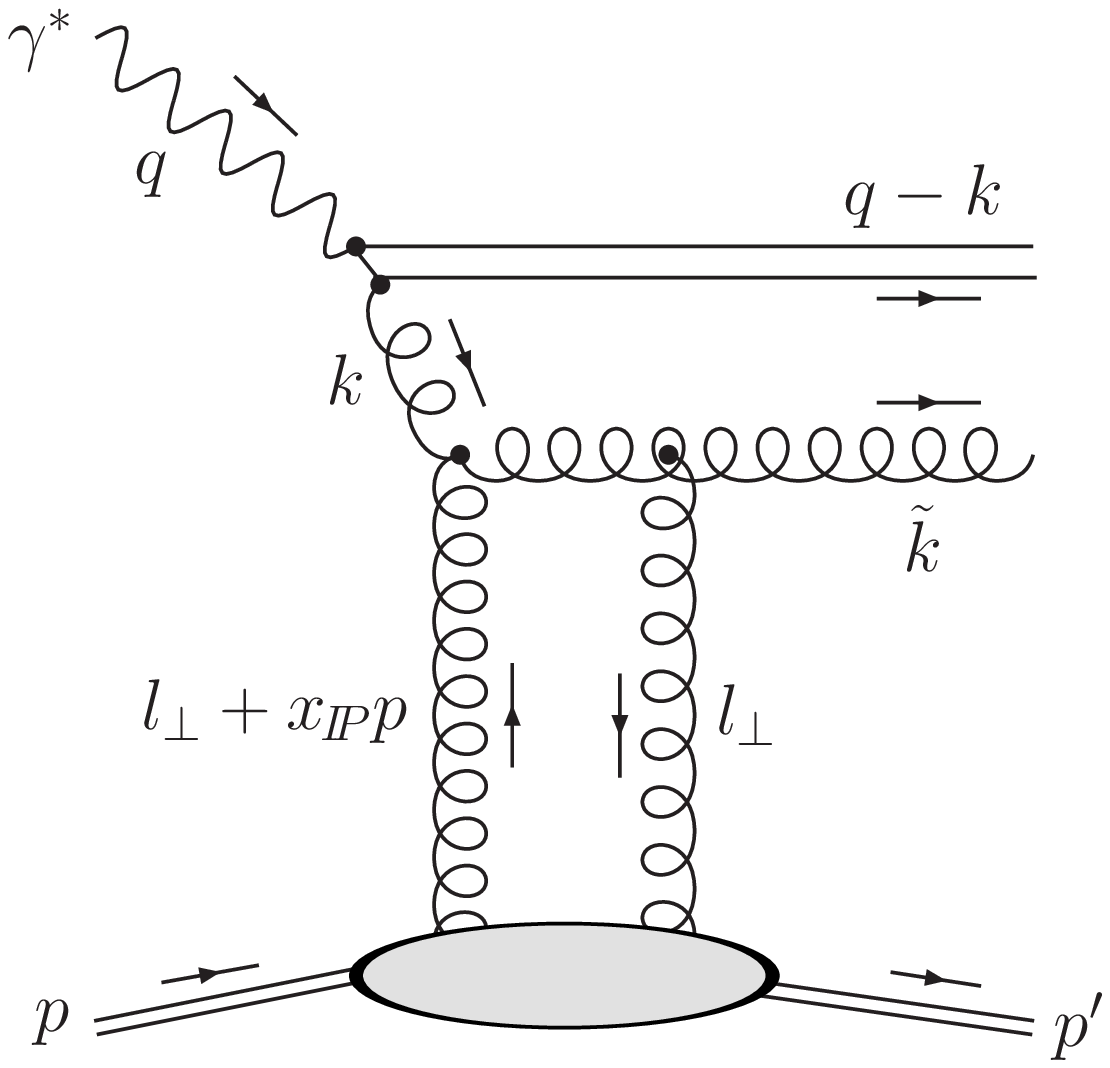}
  \end{minipage}
  \caption{(a) Quark dipole and (b) effective gluon dipole interacting with the proton via a perturbative Pomeron represented by two $t$-channel gluons.}
  \label{fig:dipoleG}
\end{figure}

First we consider the kinematics of the quark dipole shown in Fig.~\ref{fig:dipoleG}(a).  We use a Sudakov decomposition of the momentum $k$ of the off-shell quark,
\begin{equation} \label{eq:kSudakov}
  k = \alpha\,q^\prime + \beta_k\,p + k_\perp,
\end{equation}
with $q^\prime\equiv q + \xB\,p$, ${q^\prime}^2=0=p^2$, $k_\perp^2=-k_t^2$.  The two outgoing components of the dipole have momenta
\begin{gather}
  q-k = (1-\alpha)\,q^\prime + \xB\frac{(k_t^2+m_f^2)/Q^2}{1-\alpha}\,p - k_\perp,\label{eq:qminusk}\\
  \tilde{k} = k+\xPom\,p = \alpha\,q^\prime + \xB\frac{(k_t^2+m_f^2)/Q^2}{\alpha}\,p + k_\perp,\label{eq:ktilde}
\end{gather}
where the on-shell conditions, $(q-k)^2=m_f^2$ and $\tilde{k}^2=m_f^2$, determine
\begin{equation}
  \beta_k = -\xB\left(1+\frac{(k_t^2+m_f^2)/Q^2}{1-\alpha}\right),\qquad \xPom=\xB\left(1+\frac{(k_t^2+m_f^2)/Q^2}{\alpha(1-\alpha)}\right).
\end{equation}
The invariant mass of the $q\bar{q}$ system is given by
\begin{equation} \label{eq:GqqMXsq}
 M_X^2 = (q+\xPom\,p)^2 = \frac{k_t^2+m_f^2}{\alpha(1-\alpha)}.
\end{equation}
The kinematic limit occurs when $\alpha=1/2$, giving a maximum value for $k_t^2$ of $M_X^2/4-m_f^2$.  Since $\beta = Q^2/(Q^2+M_X^2)$, \eqref{eq:GqqMXsq} can be written as
\begin{equation} \label{eq:alphakt-betamu}
 \alpha(1-\alpha)Q^2 = \beta\frac{k_t^2+m_f^2}{1-\beta} \equiv \beta\mu^2,
\end{equation}
where the last equivalence specifies our choice of factorisation scale $\mu$. The off-shell quark with momentum $k$ in Fig.~\ref{fig:dipoleG}(a) has virtuality given by
\begin{equation}
  k^2-m_f^2 = -\frac{\mu^2}{1-\alpha}\simeq -\mu^2,
\end{equation}
since $\alpha\ll 1$ in the approximation of strongly-ordered transverse momentum, $\mu^2\ll Q^2$, to which we are working.\footnote{Actually, from \eqref{eq:alphakt-betamu}, $\mu^2\ll Q^2$ implies either $\alpha\ll 1$ or $(1-\alpha)\ll 1$, but it is conventional to take the former.}

\subsubsection{Quark dipole with a transversely polarised photon}

The differential $\gamma^*p$ cross section corresponding to Fig.~\ref{fig:dipoleG}(a) is given by a $k_t$-factorisation formula.  It can be written in terms of photon wave functions $\Psi(\alpha,\vec{k_t})$, describing the fluctuation of the photon into a quark--antiquark dipole, convoluted over $\alpha$ and $k_t$ with a dipole cross section $\hat{\sigma}$, describing the interaction of the dipole with the proton via two-gluon exchange.  The dipole factorisation formula for Fig.~\ref{fig:dipoleG}(a) with a transversely polarised photon is
\begin{equation} \label{eq:GqqT}
  \left.\frac{\dif\sigma_{T,q\bar{q}}^{\gamma^*p}}{\dif t}\right\rvert_{t=0} = \frac{N_C}{16\pi}\int_0^1\!\dif\alpha\int\!\frac{\dif k_t^2}{2\pi}\sum_f e_f^2\,\alpha_{\mathrm{em}}\;\frac{1}{2}\sum_{\gamma,h,h^\prime=\pm1}\left\lvert\int\!\frac{\dif^2\vec{l_t}}{\pi}\;D\Psi_{hh^\prime}^\gamma\;\frac{\dif\hat{\sigma}}{\dif l_t^2}\right\rvert^2,
\end{equation}
where the $q\bar{q}$ dipole wave functions, $\Psi_{hh^\prime}^\gamma$, and the operation $D$ are specified below. This formula for diffractive DIS may be compared to the corresponding result for inclusive DIS,\footnote{Note the extra factor 2 in \eqref{eq:GqqTDIS} compared to (7) of \cite{Wusthoff:1997fz}.}
\begin{equation} \label{eq:GqqTDIS}
  \sigma_{T,q\bar{q}}^{\gamma^*p} = N_C\int_0^1\!\dif\alpha\int\!\frac{\dif k_t^2}{2\pi}\sum_f e_f^2\,\alpha_{\mathrm{em}}\;\frac{1}{2}\sum_{\gamma,h,h^\prime=\pm1}\int\!\frac{\dif^2\vec{l_t}}{\pi}\;\left\lvert\Psi_{hh^\prime}^\gamma(\alpha,\vec{k_t})-\Psi_{hh^\prime}^\gamma(\alpha,\vec{k_t}+\vec{l_t})\right\rvert^2\;\frac{\dif\hat{\sigma}}{\dif l_t^2}.
\end{equation}
The light-cone wave functions for the quark--antiquark dipole with a transversely polarised photon are \cite{Wusthoff:1997fz,Levin:1996vf}
\begin{equation}
  \Psi_{hh^\prime}^\gamma(\alpha,\vec{k_t}) = \frac{\delta_{h,-h^\prime}\,[(1-2\alpha)h-\gamma]\,\vec{\epsilon}_\gamma\cdot\vec{k_t} + \delta_{hh^\prime}\;m_f\,h}{k_t^2+m_f^2+\alpha(1-\alpha)Q^2},
\end{equation}
where $\gamma$, $h$, and $h^\prime$ denote the helicities of the photon, quark, and antiquark, respectively.  Here, $\vec{k_t} \equiv k_t^1+\mathrm{i}\,k_t^2=(k_t^1,k_t^2)$ and the circular polarisation vectors of the photon are $\vec{\epsilon}_\gamma=(1,\gamma)/\sqrt{2}$.  The denominator of these wave functions is the virtuality of the off-shell quark with momentum $k$:
\begin{equation}
  k_t^2+m_f^2+\alpha(1-\alpha)Q^2 = (1-\beta)\mu^2 + \beta\mu^2 = \mu^2 \simeq \lvert k^2-m_f^2\rvert.
\end{equation}
Note that the wave functions are symmetric under $\alpha\to(1-\alpha)$ and $\vec{k_t}\to-\vec{k_t}$, corresponding to $q\leftrightarrow\bar{q}$, that is, $\eqref{eq:qminusk}\leftrightarrow\eqref{eq:ktilde}$, so we only need to sum over flavours in \eqref{eq:GqqT} and not over quarks and antiquarks separately.

\begin{figure}
  \centering
  \begin{minipage}{\textwidth}
    \hspace{0.1\textwidth}(a)\\
    \includegraphics[width=\textwidth]{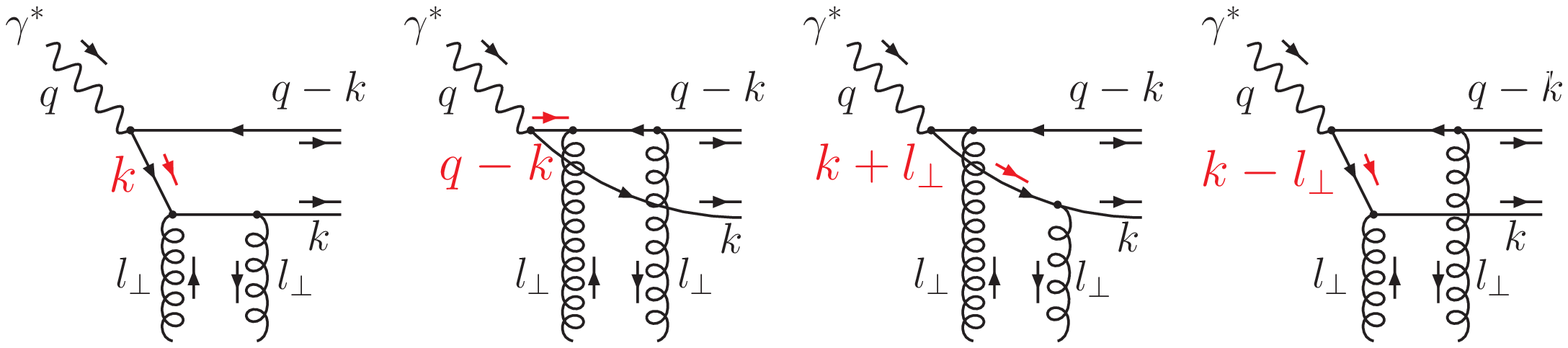}\\
  \end{minipage}
  \begin{minipage}{\textwidth}
    \hspace{0.1\textwidth}(b)\\
    \includegraphics[width=\textwidth]{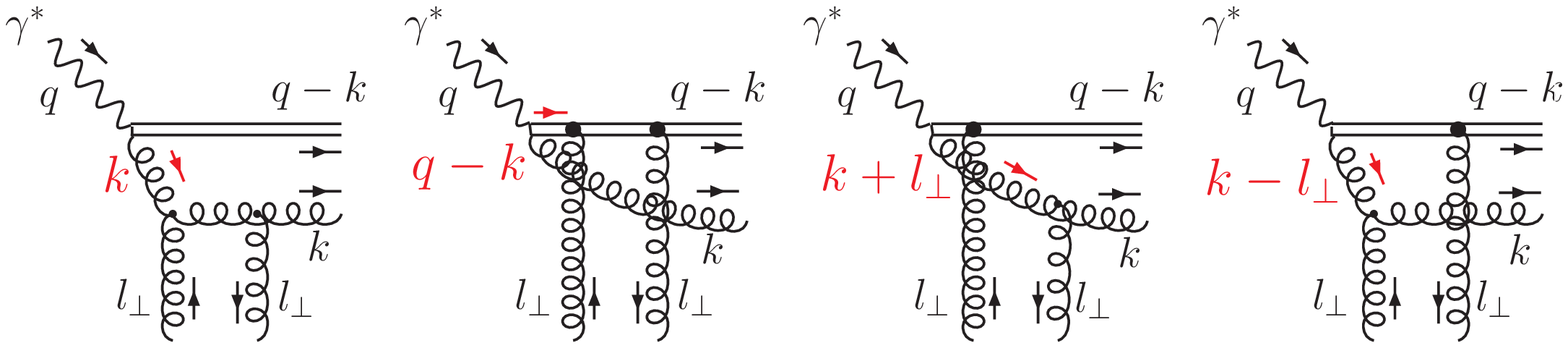}
  \end{minipage}
  \caption{The four different permutations of the couplings of the two $t$-channel gluons to the two components of (a) the quark dipole and (b) the effective gluon dipole.}
  \label{fig:permG}
\end{figure}
The four different permutations of the couplings of the two $t$-channel gluons to the two components of the quark dipole, shown in Fig.~\ref{fig:permG}(a), are obtained by simply shifting the argument of the wave function:
\begin{equation} \label{eq:DPsi}
  D\Psi(\alpha,\vec{k_t},\vec{l_t}) \equiv 2\Psi(\alpha,\vec{k_t}) - \Psi(\alpha,\vec{k_t}+\vec{l_t}) - \Psi(\alpha,\vec{k_t}-\vec{l_t}).
\end{equation}
We choose a basis where $\vec{k_t}=k_t\,(1,0)$ and $\vec{l_t}=l_t\,(\cos\phi,\sin\phi)$ and neglect the $\xPom\,p$ components of the momenta.  We work in the approximation of strongly-ordered transverse momenta, $l_t\ll \mu\ll Q$, and expand $D\Psi$ in the limit $l_t\to0$, only keeping the leading term proportional to $l_t^2$.  After performing the azimuthal integral, we find
\begin{multline}
  \int_0^{2\pi}\!\frac{\dif\phi}{2\pi}D\Psi_{hh^\prime}^\gamma(\alpha,\vec{k_t},\vec{l_t}) = \frac{l_t^2}{\left[k_t^2+m_f^2+\alpha(1-\alpha)Q^2\right]^3}\\
\times\left\{4[\alpha(1-\alpha)Q^2+m_f^2]\,\delta_{h,-h^\prime}\,[(1-2\alpha)h-\gamma]\,\vec{\epsilon}_\gamma\cdot\vec{k_t} \right.\\\left.+ 2\,\delta_{hh^\prime}\;m_f\,h\,[\alpha(1-\alpha)Q^2+m_f^2-k_t^2]\right\};
\end{multline}
cf.~(21) of \cite{Wusthoff:1997fz}.  After changing variables from $\alpha$ and $k_t^2$ to $\beta$ and $\mu^2$ using \eqref{eq:alphakt-betamu}, we obtain
\begin{multline} \label{eq:GqqTwave}
  \sum_{\gamma,h,h^\prime=\pm1}\left\lvert\int_0^{2\pi}\!\frac{\dif\phi}{2\pi}D\Psi_{hh^\prime}^\gamma(\alpha,\vec{k_t},\vec{l_t})\right\rvert^2 = l_t^4\;\frac{64}{\mu^{12}}\;\Bigg\{\left(m_f^2+\beta\mu^2\right)^2\left[(1-\beta)\mu^2-m_f^2\right]\left(1-2\beta\frac{\mu^2}{Q^2}\right)\\\left.+m_f^2\left[m_f^2+\left(\beta-\frac{1}{2}\right)\mu^2\right]^2\right\}.
\end{multline}

\begin{figure}
  \centering
  \begin{minipage}{0.5\textwidth}
    \hspace{0.35\textwidth}(a)
    \begin{center}
      \includegraphics[width=0.8\textwidth]{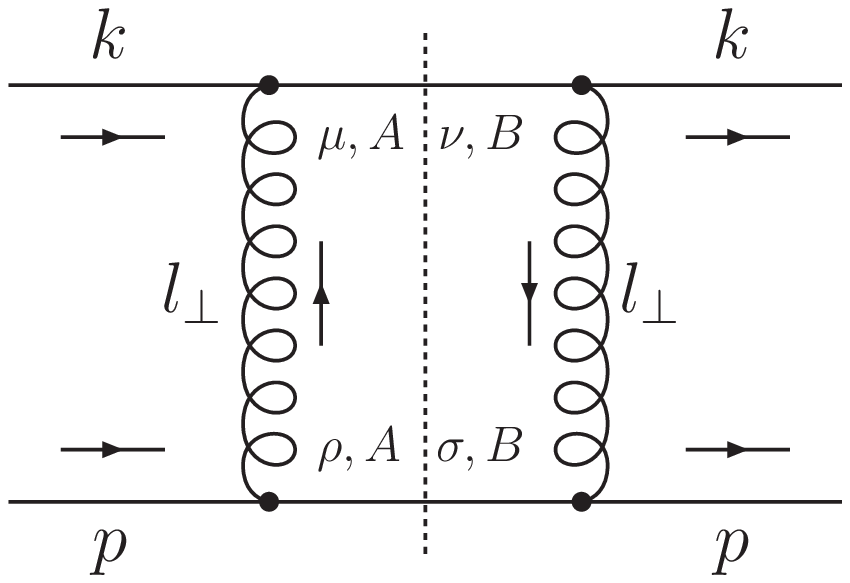}
    \end{center}
  \end{minipage}%
  \begin{minipage}{0.5\textwidth}
    \hspace{0.35\textwidth}(b)
    \begin{center}
      \includegraphics[width=0.8\textwidth]{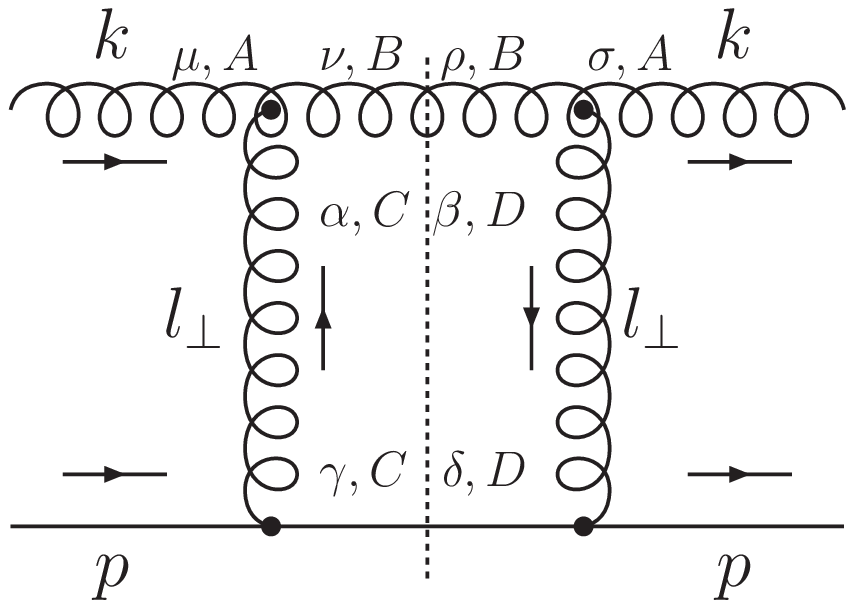}
    \end{center}
  \end{minipage}
  \caption{Cut diagrams giving the dipole cross sections for the two-gluon Pomeron: (a) $qq\to qq$ and (b) $gq\to gq$.}
  \label{fig:GdipoleX}
\end{figure}
The other necessary part of the calculation is the cross section for $qp\to qp$, which may be obtained from that for the process $qq\to qq$ with $t$-channel gluon exchange, shown in Fig.~\ref{fig:GdipoleX}(a).  We assume that $k^2\simeq 0$ and neglect the $\xPom\,p$ components of the momenta in these calculations.  The $qq\to qq$ differential cross section is
\begin{equation}
  \frac{\dif\hat\sigma}{\dif|\hat{t}|}(qq\to qq) = \frac{|\mathcal{M}|^2}{16\pi\hat{s}^2},
 \end{equation}
where $\hat{s}=2\,k\cdot p=\alpha\,Q^2/x_B\simeq\mu^2/\xPom$ and $\hat{t}=-l_t^2$.  The squared matrix element for $qq\to qq$ is
\begin{equation}
  |\mathcal{M}|^2 = \frac{1}{4}\mathcal{C}\;\frac{g^4}{l_t^4}\;\mathrm{Tr}[\gamma^\mu\slashed{k}\gamma^\nu(\slashed{k}+\slashed{l}_\perp)]\;\mathrm{Tr}[\gamma^\rho\slashed{p}\gamma^\sigma(\slashed{p}-\slashed{l}_\perp)]\;(-g_{\mu\rho})(-g_{\nu\sigma}),
\end{equation}
where the colour factor is
\begin{equation}
  \mathcal{C}(qq\to qq) = \frac{1}{N_C^2}\mathrm{Tr}[t^At^B]\mathrm{Tr}[t^At^B] = \frac{2}{9}.
\end{equation}
In the high-energy limit, where only the terms to leading $\mathcal{O}(|\hat{t}|/\hat{s})$ are retained, that is, in the $l_t\to 0$ limit, we have
\begin{equation}
  \frac{\dif\hat\sigma}{\dif l_t^2}(qq\to qq) = \frac{1}{l_t^4}\frac{8\pi}{9}\alpha_S(l_t^2)\alpha_S(\mu^2),
\end{equation}
where appropriate scales for $\alpha_S$ have been chosen corresponding to the two different vertices in Fig.~\ref{fig:GdipoleX}(a).  The lower vertex in Fig.~\ref{fig:GdipoleX}(a) may be considered as the first step of DGLAP evolution, which generates the unintegrated gluon distribution of the proton, $f_g(\xPom,l_t^2,\mu^2)$.  Therefore, we obtain the cross section for $qp\to qp$ by making the replacement
\begin{equation} \label{eq:Pgq-fg}
  \left.\frac{\alpha_S(l_t^2)}{2\pi}\xPom P_{gq}(\xPom)\right\rvert_{\xPom\ll 1} = \frac{\alpha_S(l_t^2)}{2\pi}2C_F=\frac{4}{3\pi}\alpha_S(l_t^2)\to f_g(\xPom,l_t^2,\mu^2).
\end{equation}
This replacement accounts for more complicated diagrams than Fig.~\ref{fig:GdipoleX}(a) which include the complete DGLAP evolution, leading to
\begin{equation} \label{eq:Gqpqp}
  \frac{\dif\hat\sigma}{\dif l_t^2}(qp\to qp) = \frac{1}{l_t^4}\frac{2\pi^2}{3}\alpha_S(\mu^2)f_g(\xPom,l_t^2,\mu^2).
\end{equation}
Combining \eqref{eq:GqqTwave} and \eqref{eq:Gqpqp} we obtain
\begin{align}
  \sum_{\gamma,h,h^\prime=\pm1}\left\lvert\int\!\frac{\dif^2\vec{l_t}}{\pi}\;D\Psi_{hh^\prime}^\gamma\;\frac{\dif\hat{\sigma}}{\dif l_t^2}\right\rvert^2 &= \frac{64}{\mu^{12}}\left\{\ldots\right\}\left[\frac{2\pi^2}{3}\alpha_S(\mu^2)\int_0^{\mu^2}\!\diff{l_t^2}f_g(\xPom,l_t^2,\mu^2)\right]^2 \notag \\&=\frac{256\pi^4}{9\mu^{12}}\left[\,\alpha_S(\mu^2)\;\xPom g(\xPom,\mu^2)\,\right]^2\left\{\ldots\right\}, \label{eq:GPsiSigHat}
\end{align}
where $g(\xPom,\mu^2)$ is the \emph{integrated} gluon distribution of the proton and $\left\{\ldots\right\}$ denotes everything inside the curly brackets in \eqref{eq:GqqTwave}.  Strictly speaking, this last expression should be written in terms of the off-diagonal  (or skewed) gluon distribution of the proton, since the left and right $t$-channel gluons in Fig.~\ref{fig:dipoleG}(a) carry different fractions of the proton momentum.   At small $\xPom$, and assuming that $\xPom\,g(\xPom,\mu^2)\propto \xPom^{-\lambda}$, then the off-diagonal gluon distribution is given by the diagonal distribution multiplied by an overall constant factor \cite{Shuvaev:1999ce},\footnote{This factor is not seen in calculating Fig.~\ref{fig:GdipoleX}(a) since it is absent in the limit $\xPom\to 0$ \cite{Bartels:1981jh}.}
\begin{equation} \label{eq:Rg}
  R_g(\lambda) = \frac{2^{2\lambda+3}}{\sqrt{\pi}}\frac{\Gamma(\lambda+5/2)}{\Gamma(\lambda+4)}.
\end{equation}
Changing the integration variables in \eqref{eq:GqqT},
\begin{equation}
  \dif\alpha\;\dif k_t^2 = \dif\beta\;\dif\mu^2\;\frac{\mu^2/Q^2}{\sqrt{1-4\beta\mu^2/Q^2}},
\end{equation}
then integrating \eqref{eq:GqqT} over $t$, assuming a $t$ dependence of the form $\exp(B_D\,t)$, gives
\begin{equation} \label{eq:GSigT}
  \sigma_{T,q\bar{q}}^{\gamma^*p} = \frac{R_g^2}{B_D}\frac{N_C}{16\pi}\int_0^1\!\dif\beta\int\!\frac{\dif\mu^2}{2\pi}\frac{\mu^2/Q^2}{\sqrt{1-4\beta\mu^2/Q^2}}\sum_f e_f^2\,\alpha_{\mathrm{em}}\;\frac{1}{2}\sum_{\gamma,h,h^\prime=\pm1}\left\lvert\int\!\frac{\dif^2\vec{l_t}}{\pi}\;D\Psi_{hh^\prime}^\gamma\;\frac{\dif\hat{\sigma}}{\dif l_t^2}\right\rvert^2.
\end{equation}
The relation between the diffractive structure function and the $\gamma^* p$ cross section is 
\begin{equation} \label{eq:f2d3gammap}
  \fd = \frac{Q^2}{4\pi^2\alpha_{\mathrm{em}}}\frac{\beta}{\xPom}\frac{\dif\sigma^{\gamma^*p}}{\dif\beta}.
\end{equation}
Thus, on inserting \eqref{eq:GPsiSigHat} into \eqref{eq:GSigT} we finally obtain
\begin{equation} \label{eq:ftd3mass}
  F_{T,q\bar{q}}^{{\rm D}(3)} = \sum_fe_f^2\int_{\frac{m_f^2}{1-\beta}}^{\frac{Q^2}{4\beta}}\!\diff{\mu^2}\;\frac{1}{\sqrt{1-4\beta\mu^2/Q^2}}\frac{1}{\xPom B_D} \left[\,R_g\frac{\alpha_S(\mu^2)}{\mu}\;\xPom g(\xPom,\mu^2)\,\right]^2\;\frac{\beta}{3\mu^6}\left\{\ldots\right\},
\end{equation}
where again $\left\{\ldots\right\}$ denotes everything inside the curly brackets in \eqref{eq:GqqTwave}.  Replacing $\sum_fe_f^2$ by $e_c^2$, this equation corresponds to diffractive open charm production; cf.~(45) of \cite{Levin:1996vf}.

We now take the limit of massless quarks, $m_f\to0$ where $f=u,d,s$, in the approximation where $\mu^2\ll Q^2$.  In this approximation, we can replace the upper limit of the $\mu^2$ integral, $Q^2/(4\beta)$, by $Q^2$.  We also need to replace the lower limit by an infrared cutoff $\mu_0^2$.  Then \eqref{eq:ftd3mass} becomes
\begin{equation} \label{eq:ftd3}
  F_{T,q\bar{q}}^{{\rm D}(3)} = \int_{\mu_0^2}^{Q^2}\!\diff{\mu^2}\;\frac{1}{\xPom B_D} \left[\,R_g\frac{\alpha_S(\mu^2)}{\mu}\;\xPom g(\xPom,\mu^2)\,\right]^2\;\langle e_f^2\rangle\beta^3(1-\beta),
\end{equation}
where $\langle e_f^2\rangle\equiv (\sum_f e_f^2)/n_f$ (with $n_f=3$), which coincides with (22) of \cite{Wusthoff:1997fz}.  We can write this expression as
\begin{equation} \label{eq:f2d3GdipoleT}
  F_{T,q\bar{q}}^{{\rm D}(3)} = \int_{\mu_0^2}^{Q^2}\!\diff{\mu^2}\;f_{\Pom=G}(\xPom;\mu^2)\;F_T^{\Pom=G}(\beta,\mu^2;\mu^2),
\end{equation}
where the `Pomeron flux factor' is\footnote{Note that this definition differs slightly from that given in \cite{MRW1,MRW2}.  Since only the combination \eqref{eq:f2d3GdipoleT} matters, we are free to redistribute factors of $\mu$ and constants as we please.}
\begin{equation} \label{eq:f2d3GdipoleTPomflux}
  f_{\Pom=G}(\xPom;\mu^2) = \frac{1}{\xPom B_D} \left[\,R_g\frac{\alpha_S(\mu^2)}{\mu}\;\xPom g(\xPom,\mu^2)\,\right]^2,
\end{equation}
and the `Pomeron structure function' at a scale $\mu$ originating from a component of the Pomeron of size $1/\mu$ is
\begin{equation} \label{eq:f2d3GdipoleTSigmaPom}
  F_T^{\Pom=G}(\beta,\mu^2;\mu^2) = \langle e_f^2\rangle\;\beta\Sigma^{\Pom=G}(\beta,\mu^2;\mu^2) = \langle e_f^2\rangle\;\beta^3(1-\beta).
\end{equation}
Recall that the notation $\Pom=G$ is used to indicate that the perturbative Pomeron is represented by two $t$-channel gluons.  The quark singlet density \eqref{eq:f2d3GdipoleTSigmaPom} can be taken as the initial condition at a scale $\mu^2$, together with the corresponding gluon density \eqref{eq:gG}, for DGLAP evolution up to $Q^2$, giving $\Sigma^{\Pom=G}(\beta,Q^2;\mu^2)$ and $g^{\Pom=G}(\beta,Q^2;\mu^2)$; see Fig.~\ref{fig:DDISladder}.  In this context, $\beta\Sigma^{\Pom=G}(\beta,\mu^2;\mu^2)$ can be regarded as the Pomeron-to-quark splitting function, $P_{q,\Pom=G}(\beta)$; see \eqref{eq:spfn}.

\subsubsection{Quark dipole with a longitudinally polarised photon}
The dipole factorisation formula for Fig.~\ref{fig:dipoleG}(a) with a longitudinally polarised photon is
\begin{equation}
  \left.\frac{\dif\sigma_{L,q\bar{q}}^{\gamma^*p}}{\dif t}\right\rvert_{t=0} = \frac{N_C}{16\pi}\int_0^1\!\dif\alpha\int\!\frac{\dif k_t^2}{2\pi}\sum_f e_f^2\,\alpha_{\mathrm{em}}\;\sum_{h,h^\prime=\pm1}\left\lvert\int\!\frac{\dif^2\vec{l_t}}{\pi}\;D\Psi_{hh^\prime}^{\gamma=0}\;\frac{\dif\hat{\sigma}}{\dif l_t^2}\right\rvert^2,
\end{equation}
where the light-cone wave functions for the quark-antiquark dipole with a longitudinally polarised photon are \cite{Wusthoff:1997fz,Levin:1996vf}
\begin{equation}
  \Psi_{hh^\prime}^{\gamma=0}(\alpha,\vec{k_t}) = \frac{2\alpha(1-\alpha)Q}{k_t^2+m_f^2+\alpha(1-\alpha)Q^2}\;\delta_{h,-h^\prime}.
\end{equation}
Taking the limit $l_t\to0$ of the combination of wave functions \eqref{eq:DPsi} and performing the azimuthal integral gives\footnote{Note the extra factor 2 in \eqref{eq:GqqLlt0} compared to (21) of \cite{Wusthoff:1997fz}.}
\begin{equation} \label{eq:GqqLlt0}
  \int_0^{2\pi}\!\frac{\dif\phi}{2\pi}D\Psi_{hh^\prime}^{\gamma=0}(\alpha,\vec{k_t},\vec{l_t}) = l_t^2 \frac{\alpha(1-\alpha)Q^2+m_f^2-k_t^2}{\left[k_t^2+m_f^2+\alpha(1-\alpha)Q^2\right]^3}\;4\alpha(1-\alpha)Q\;\delta_{h,-h^\prime}.
\end{equation}
Squaring this expression and summing over the quark helicities gives
\begin{equation}
  \sum_{h,h^\prime=\pm1} \left\lvert\int_0^{2\pi}\!\frac{\dif\phi}{2\pi}D\Psi_{hh^\prime}^{\gamma=0}(\alpha,\vec{k_t},\vec{l_t})\right\rvert^2 = l_t^4\frac{32}{\mu^6}\frac{\mu^2}{Q^2}\beta^2\left(2\beta-1+\frac{2m_f^2}{\mu^2}\right)^2.
\end{equation}
The dipole cross section, $\dif\hat\sigma/\dif l_t^2(qp\to qp)$, is the same as \eqref{eq:Gqpqp}, giving
\begin{equation} \label{eq:fld3mass}
  F_{L,q\bar{q}}^{{\rm D}(3)} = \sum_f e_f^2\int_{\frac{m_f^2}{1-\beta}}^{\frac{Q^2}{4\beta}}\!\diff{\mu^2}\;\frac{1}{\sqrt{1-4\beta\mu^2/Q^2}}\frac{\mu^2}{Q^2}\frac{1}{\xPom B_D} \left[\,R_g\frac{\alpha_S(\mu^2)}{\mu}\;\xPom g(\xPom,\mu^2)\,\right]^2\;\frac{\beta^3}{3}\left(2\beta-1+\frac{2m_f^2}{\mu^2}\right)^2,
\end{equation}
which corresponds to diffractive open charm production, cf.~(44) of \cite{Levin:1996vf}, after replacing $\sum_fe_f^2$ by $e_c^2$.  As before we take the limit of massless quarks in the approximation where $\mu^2\ll Q^2$, then
\begin{equation} \label{eq:LG}
  F_{L,q\bar{q}}^{{\rm D}(3)} = \frac{Q^2}{4\pi^2\alpha_{\mathrm{em}}}\frac{\beta}{\xPom}\frac{\dif\sigma_{L,q\bar{q}}^{\gamma^*p}}{\dif\beta} = \left(\int_{\mu_0^2}^{Q^2}\!\diff{\mu^2}\;\frac{\mu^2}{Q^2}\;f_{\Pom=G}(\xPom;\mu^2)\right)\;F_L^{\Pom=G}(\beta),
\end{equation}
where $F_L^{\Pom=G}(\beta) = \langle e^2_f\rangle\;\beta^3\left(2\beta-1\right)^2$; this result is a factor 2 different from (23) of \cite{Wusthoff:1997fz}, but is in agreement with \cite{Levin:1996vf}.  Note that this contribution to $\fd$ is twist-four due to the extra factor $\mu^2/Q^2$ with respect to \eqref{eq:f2d3GdipoleT}.  Here we include only this main extra $1/Q^2$ dependence, and omit the $\mathcal{O}(\alpha_S)$ anomalous dimension arising from twist-four DGLAP evolution.

\subsubsection{Gluon dipole with a transversely polarised photon}

Now consider the kinematics of the $q\bar{q}g$ system shown in Fig.~\ref{fig:dipoleG}(b).  Although this diagram has an extra factor $\alpha_S$ with respect to the $q\bar{q}$ system shown in Fig.~\ref{fig:dipoleG}(a), it is known to be dominant at large $M_X$ (small $\beta$) due to $t$-channel spin-1 gluon exchange.  Using a Sudakov parameterisation \eqref{eq:kSudakov} of the momentum $k$ of the off-shell gluon gives the momenta of the outgoing $q\bar{q}$ pair and gluon as
\begin{gather} \label{eq:j1}
  q-k = (1-\alpha)\,q^\prime + \frac{\xB}{Q^2}\frac{k_t^2+M_{q\bar{q}}^2}{1-\alpha}\,p - k_\perp,\\
  \tilde{k} = k+\xPom\,p = \alpha\,q^\prime + \xB\frac{k_t^2/Q^2}{\alpha}\,p + k_\perp,   \label{eq:j2}
\end{gather}
respectively, where $M_{q\bar{q}}$ is the invariant mass of the $q\bar{q}$ system.  Again the on-shell conditions, $(q-k)^2=M_{q\bar{q}}^2$ and $\tilde{k}^2=0$, determine
\begin{equation}  \label{eq:j3}
  \beta_k = -\xB\left(1+\frac{k_t^2+M_{q\bar{q}}^2}{(1-\alpha)Q^2}\right),\qquad \xPom=\xB\left(1+\frac{k_t^2+\alpha M_{q\bar{q}}^2}{\alpha(1-\alpha)Q^2}\right).
\end{equation}
The invariant mass of the $q\bar{q}g$ system is given by
\begin{equation}  \label{eq:j4}
 M_X^2 = (q+\xPom\,p)^2 = \frac{k_t^2+\alpha M_{q\bar{q}}^2}{\alpha(1-\alpha)}.
\end{equation}
The maximum value of $k_t^2$ occurs when $\alpha=(1-M_{q\bar{q}}^2/M_X^2)/2$, giving an upper limit for $k_t^2$ of
\begin{equation}
  k_t^2 = \frac{M_X^2}{4}\left(1-\frac{M_{q\bar{q}}^2}{M_X^2}\right)^2.
\end{equation}
Defining the (light-cone) fraction of the Pomeron's momentum carried by the gluon with momentum $\tilde{k}$ in Fig.~\ref{fig:dipoleG}(b) to be
\begin{equation}
  \tilde{\beta} = \left(\xB\frac{k_t^2/Q^2}{\alpha}\right)/\xPom,
\end{equation}
then the off-shell gluon with momentum $k$ in Fig.~\ref{fig:dipoleG}(b) has virtuality
\begin{equation} \label{eq:ladderVirtuality}
  k^2  = -k_t^2/\tilde{\beta} \equiv -\mu^2.
\end{equation}
Referring to Fig.~\ref{fig:DDISladder}, the same kinematics hold if we now replace $M_{q\bar{q}}$ by the invariant mass of all the emitted partons above the one labelled $\tilde{k}$ in Fig.~\ref{fig:DDISladder}.

The $q\bar{q}g$ calculation is greatly simplified in the approximation $M_{q\bar{q}}\ll Q$, in addition to assuming strongly-ordered transverse momenta, $l_t\ll k_t\ll Q$.  In this limit, the kinematics of the $(q\bar{q})g$ system is identical to the previously considered $q\bar{q}$ system of Fig.~\ref{fig:dipoleG}(a) in the case of massless quarks.  The emitted $(q\bar{q})$ pair is localised in impact parameter space, and forms an effective `gluon' conjugate in colour to the emitted gluon.  The $(q\bar{q})g$ system can thus be considered as forming an effective $gg$ dipole.  This argument generalises if the $(q\bar{q})$ pair is replaced by the emitted partons above the one labelled $\tilde{k}$ in Fig.~\ref{fig:DDISladder}, all of which are strongly ordered in transverse momentum (see \cite{Ryskin:1990fb,Dokshitzer:1977sg}).

The dipole factorisation formula for Fig.~\ref{fig:dipoleG}(b) with a transversely polarised photon is then
\begin{equation} \label{eq:GggT}
  \left.\frac{\dif\sigma_{T,gg}^{\gamma^*p}}{\dif t}\right\rvert_{t=0} = \frac{N_C^2-1}{16\pi}\int_0^1\!\dif\alpha\int\!\frac{\dif k_t^2}{2\pi}\sum_f e_f^2\,\alpha_{\mathrm{em}}\sum_{m,n=1,2}\left\lvert\int\!\frac{\dif^2\vec{l_t}}{\pi}\;D\Psi^{mn}\;\frac{\dif\hat{\sigma}}{\dif l_t^2}\right\rvert^2.
\end{equation}
Again there are four different permutations of the couplings of the two gluons to the two components of the effective gluon dipole, shown in Fig.~\ref{fig:permG}(b), which are obtained by shifting the argument of the wave function as in \eqref{eq:DPsi}.  The light-cone wave functions for the effective gluon dipole with a transversely polarised photon are \cite{Wusthoff:1997fz}
\begin{equation}
  \Psi^{mn}(\alpha,\vec{k_t}) = \frac{1}{\sqrt{\alpha(1-\alpha)Q^2}}\;\frac{k_t^2\delta^{mn}-2k_t^mk_t^n}{k_t^2+\alpha(1-\alpha)Q^2}\quad:\quad m,n=1,2.
\end{equation}
Taking the limit $l_t\to0$ of $D\Psi^{mn}$ and performing the azimuthal integral gives
\begin{equation}
  \int_0^{2\pi}\!\frac{\dif\phi}{2\pi}D\Psi^{mn}(\alpha,\vec{k_t},\vec{l_t}) = l_t^2 \frac{2k_t^2}{\sqrt{\alpha(1-\alpha)Q^2}}\frac{3\alpha(1-\alpha)Q^2+k_t^2}{\left[k_t^2+\alpha(1-\alpha)Q^2\right]^3}\left(\delta^{mn}-\frac{2k_t^mk_t^n}{k_t^2}\right);
\end{equation}
cf.~(24) of \cite{Wusthoff:1997fz}.  Squaring this expression and summing over the indices $m,n=1,2$ gives
\begin{equation} \label{eq:GggTwave}
  \sum_{m,n=1,2}\left\lvert \int_0^{2\pi}\!\frac{\dif\phi}{2\pi}D\Psi^{mn}(\alpha,\vec{k_t},\vec{l_t}) \right\rvert^2 = l_t^4\frac{8}{\mu^6}(1-\beta)^2(1+2\beta)^2\frac{1}{\beta}.
\end{equation}

The dipole cross section for $gp\to gp$ is obtained from the scattering process $gq\to gq$ with $t$-channel gluon exchange, shown in Fig.~\ref{fig:GdipoleX}(b).  Here, the squared matrix element is
\begin{multline}
  |\mathcal{M}|^2 = \frac{1}{4}\mathcal{C}\;\frac{g^4}{l_t^4}\;\mathrm{Tr}[\slashed{p}\gamma^\gamma(\slashed{p}-\slashed{l}_\perp)\gamma^\delta]\;d_{\mu\sigma}(k,p)\;d_{\nu\rho}(k+l_\perp,p)\;(-g_{\alpha\gamma})\;(-g_{\beta\delta})\\\times \left[(2k+l_\perp)^\alpha g^{\mu\nu}-(k+2l_\perp)^\mu g^{\nu\alpha}+(l_\perp-k)^\nu g^{\mu\alpha}\right] \\\times \left[(2k+l_\perp)^\beta g^{\rho\sigma}-(k+2l_\perp)^\sigma g^{\rho\beta}+(l_\perp-k)^\rho g^{\sigma\beta}\right],
\end{multline}
where the transverse polarisations of the incoming and outgoing gluons are summed in a light-cone gauge,
\begin{equation}
  d_{\mu\sigma}(k,p) \equiv -g_{\mu\sigma} + \frac{k_\mu p_\sigma + p_\mu k_\sigma}{p\cdot k},
\end{equation}
and where the colour factor is
\begin{equation}
  \mathcal{C}(gq\to gq) = \frac{1}{N_C}\frac{1}{(N_C^2-1)}\;f^{ABC}f^{ABD}\;\mathrm{Tr}[t^Ct^D] = \frac{1}{2}.
\end{equation}
In the high-energy limit, where only the terms to leading $\mathcal{O}(\hat{s}/|\hat{t}|)$ are retained, that is, in the $l_t\to0$ limit,
\begin{equation}
  \frac{\dif\hat\sigma}{\dif l_t^2}(gq\to gq) = \frac{1}{l_t^4}2\pi\alpha_S(l_t^2)\alpha_S(\mu^2),
\end{equation}
where appropriate scales for $\alpha_S$ have been chosen corresponding to the two different vertices in Fig.~\ref{fig:GdipoleX}(b).  As before, we obtain the cross section for $gp\to gp$ by making the replacement \eqref{eq:Pgq-fg}, which gives
\begin{equation} \label{eq:Ggpgp}
  \frac{\dif\hat\sigma}{\dif l_t^2}(gp\to gp) = \frac{1}{l_t^4}\frac{3\pi^2}{2}\alpha_S(\mu^2)f_g(\xPom,l_t^2,\mu^2).
\end{equation}
Inserting \eqref{eq:GggTwave} and \eqref{eq:Ggpgp} into \eqref{eq:GggT}, and accounting for the skewed effect, the change of integration variables from $(\alpha,k_t^2)$ to $(\beta,\mu^2)$, and the $t$ dependence, we obtain
\begin{equation}
  \frac{\dif\sigma_{T,gg}^{\gamma^*p}}{\dif\beta} = \frac{4\pi^2\alpha_{\mathrm{em}}}{Q^2}\int_{\mu_0^2}^{Q^2}\!\diff{\mu^2}\;\frac{1}{B_D}\left[\,R_g\frac{\alpha_S(\mu^2)}{\mu}\;\xPom g(\xPom,\mu^2)\,\right]^2\;\sum_fe_f^2\frac{9}{8}(1-\beta)^2(1+2\beta)^2\frac{1}{\beta}.
\end{equation}

We must account for the fact that the off-shell gluon with momentum $k$ in Fig.~\ref{fig:dipoleG}(b) does not interact directly with the photon, but first splits into a quark--antiquark pair forming the effective `gluon' of the dipole.  To do this, we replace $\beta\to\beta^\prime$ in the previous formula and include the DGLAP splitting for $g\to q\bar{q}$, that is,
\begin{equation} \label{eq:gtoqqbar}
  \frac{\dif\sigma_{T,(q\bar{q})g}^{\gamma^*p}}{\dif\beta} = \frac{\alpha_S(Q^2)}{2\pi}\ln\left(\frac{Q^2}{\mu^2}\right)\int_\beta^1\!\diff{\beta^\prime}P_{qg}\left(\frac{\beta}{\beta^\prime}\right)\;\frac{\dif\sigma_{T,gg}^{\gamma^*p}}{\dif\beta^\prime}.
\end{equation}
Putting everything together, we finally obtain
\begin{equation}
  F_{T,(q\bar{q})g}^{{\rm D}(3)} = \frac{Q^2}{4\pi^2\alpha_{\mathrm{em}}}\frac{\beta}{\xPom}\frac{\dif\sigma_{T,(q\bar{q})g}^{\gamma^*p}}{\dif\beta} = \int_{\mu_0^2}^{Q^2}\!\diff{\mu^2}\;f_{\Pom=G}(\xPom;\mu^2)\;F_T^{\Pom=G}(\beta,Q^2;\mu^2),
\end{equation}
where
\begin{align}
  F_T^{\Pom=G}(\beta,Q^2;\mu^2) &= \langle e_f^2\rangle\beta\Sigma^{\Pom=G}(\beta,Q^2;\mu^2) \notag\\
  &= 2\sum_fe_f^2\frac{\alpha_S(Q^2)}{2\pi}\ln\left(\frac{Q^2}{\mu^2}\right)\beta\int_\beta^1\!\frac{\dif{\beta^\prime}}{{\beta^\prime}^2}P_{qg}\left(\frac{\beta}{\beta^\prime}\right)\,\beta^\prime g^{\Pom=G}(\beta^\prime,\mu^2;\mu^2),
\end{align}
with
\begin{equation} \label{eq:gG}
  \beta^\prime g^{\Pom=G}(\beta^\prime,\mu^2;\mu^2) = \frac{9}{16}\;(1-\beta^\prime)^2(1+2\beta^\prime)^2.
\end{equation}
Note the extra factor 2 compared to (25) of \cite{Wusthoff:1997fz} (this was corrected in a later paper \cite{Golec-Biernat:1999qd}).  We take \eqref{eq:gG} at a scale $\mu^2$, which can be interpreted as the Pomeron-to-gluon splitting function $P_{g,\Pom=G}$ \eqref{eq:spfn2}, as the initial condition for DGLAP evolution up to $Q^2$.

\subsection{Two-quark exchange ($\Pom=S$)} \label{sec:two-quark-exchange}

\begin{figure}
  \centering
  \begin{minipage}{0.49\textwidth}
    \hspace{0.5\textwidth}(a)\\
    \includegraphics[width=\textwidth]{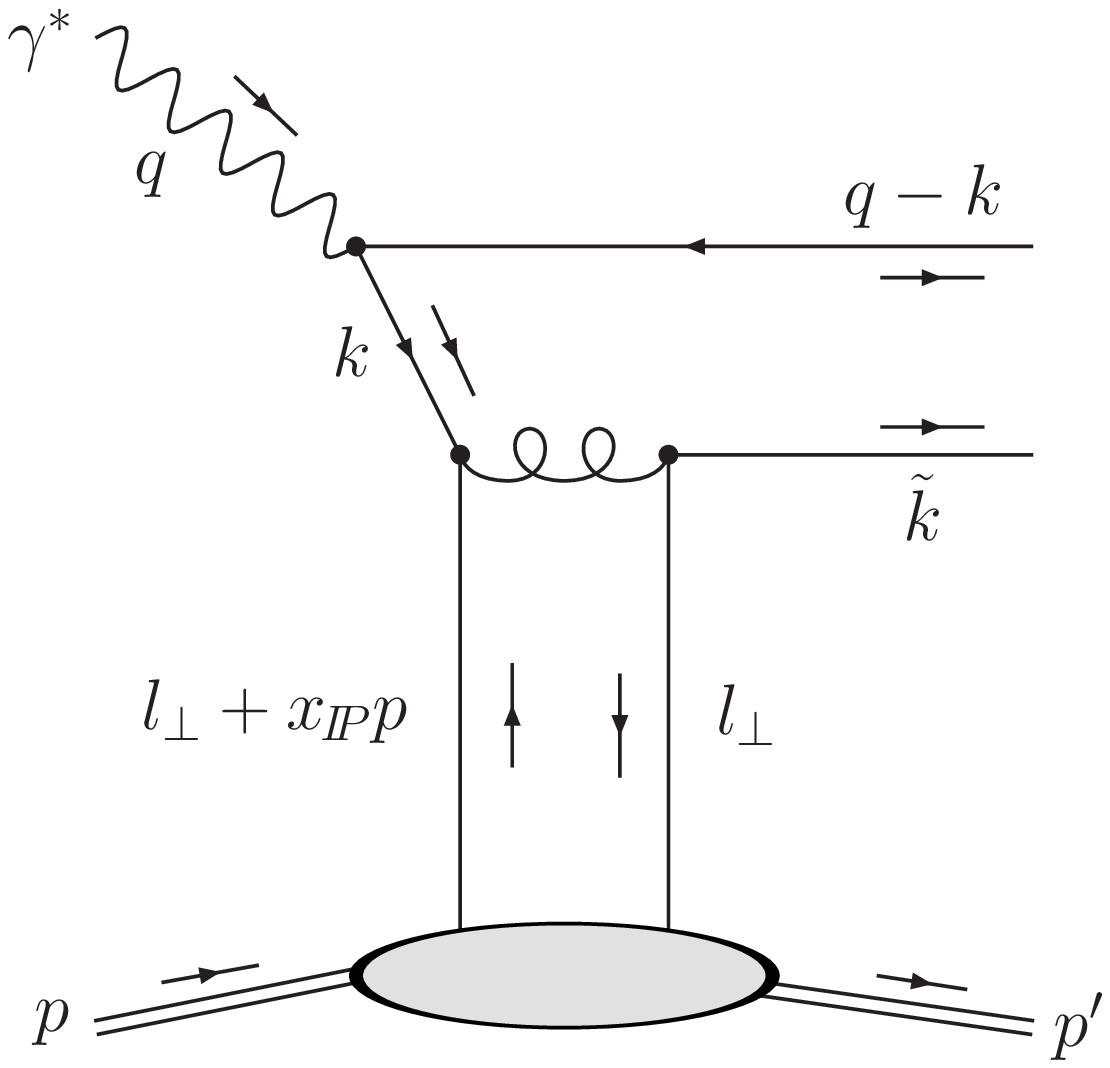}
  \end{minipage}
  \begin{minipage}{0.49\textwidth}
    \hspace{0.5\textwidth}(b)\\
    \includegraphics[width=\textwidth]{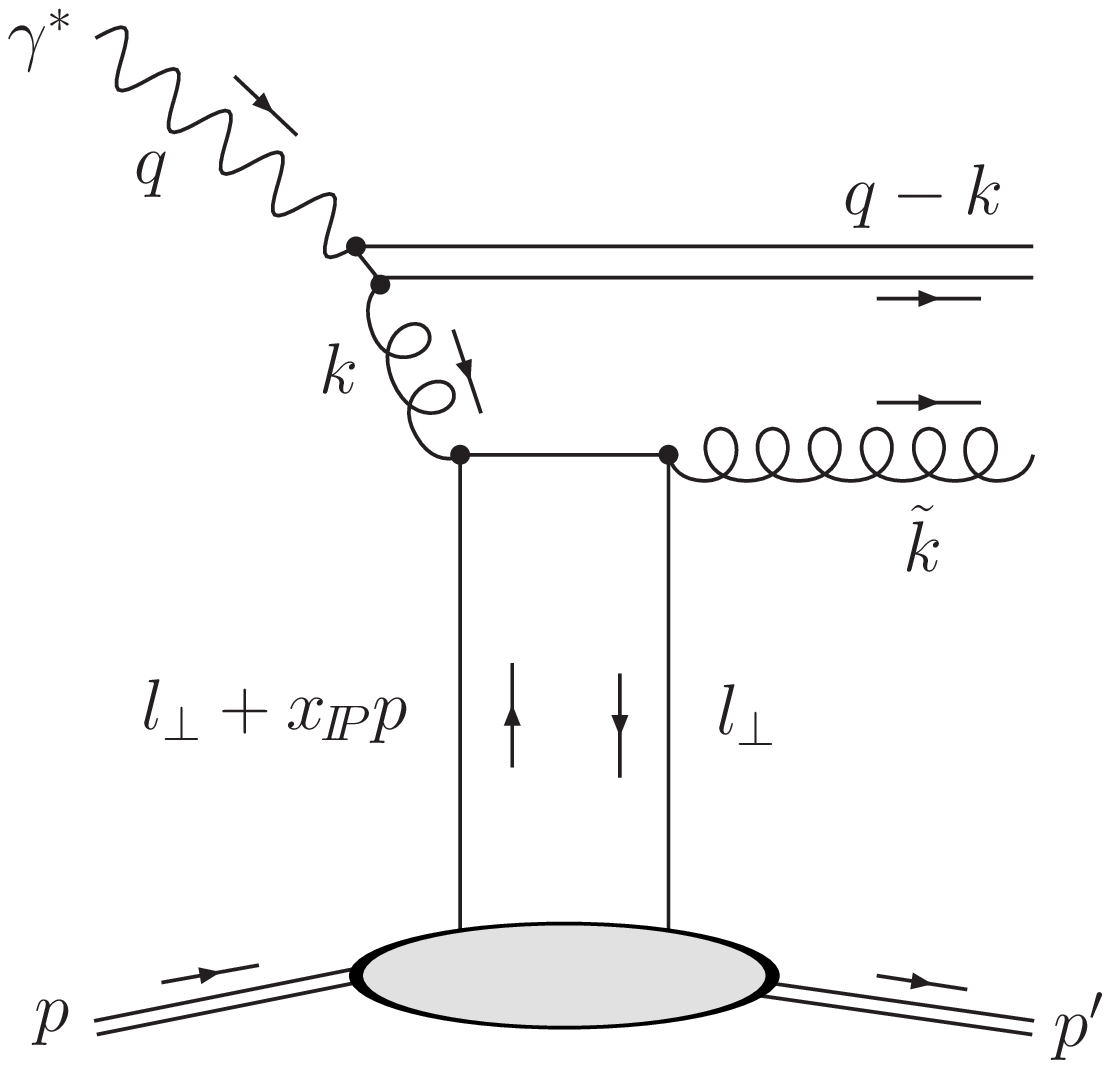}
  \end{minipage}
  \caption{(a) Quark dipole and (b) effective gluon dipole interacting with the proton via a perturbative Pomeron represented by two $t$-channel sea quarks.}
  \label{fig:dipoleS}
\end{figure}

We now calculate the lowest-order Feynman diagrams for the case in which Pomeron exchange is represented by a $t$-channel sea-quark--antiquark pair, rather than two gluons.
\begin{figure}
  \centering
  \begin{minipage}{0.7\textwidth}
    \hspace{0.2\textwidth}(a)\\
      \includegraphics[width=\textwidth]{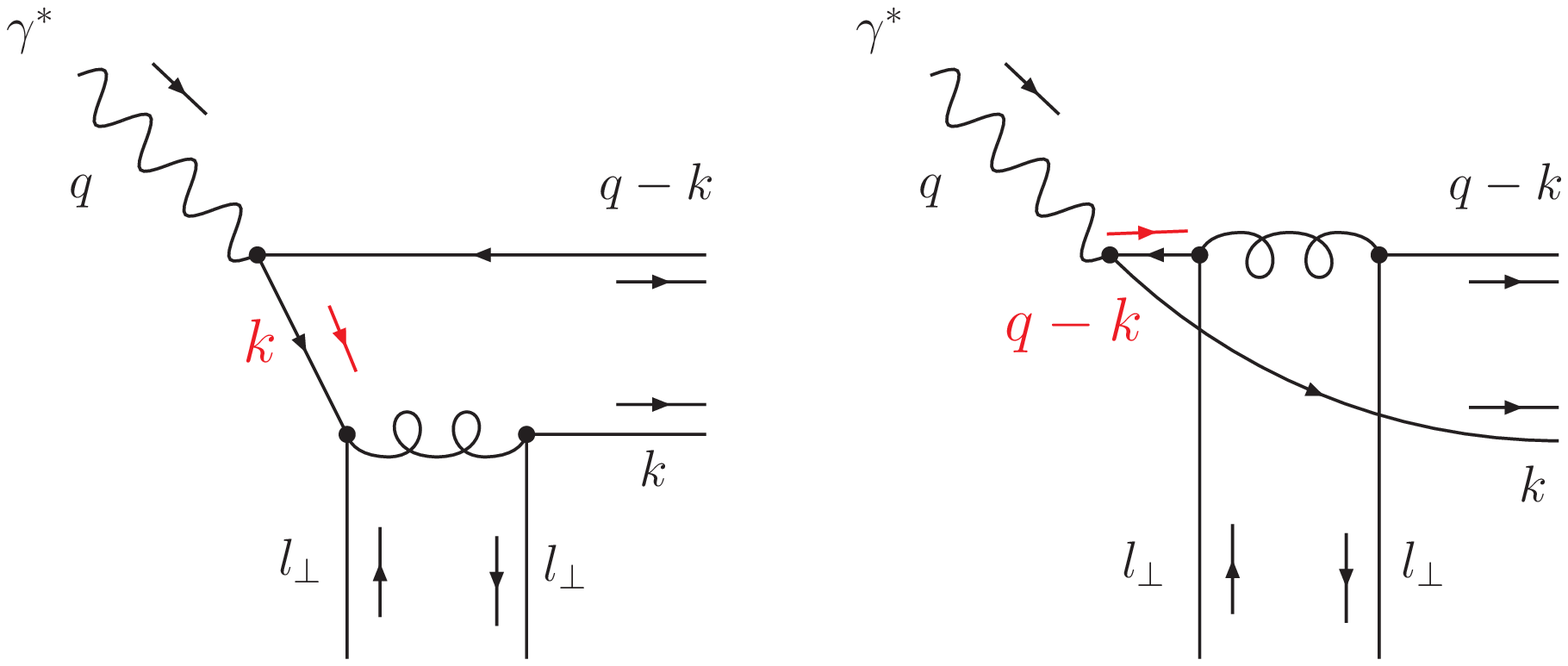}\\
  \end{minipage}
  \begin{minipage}{0.7\textwidth}
    \hspace{0.2\textwidth}(b)\\
      \includegraphics[width=\textwidth]{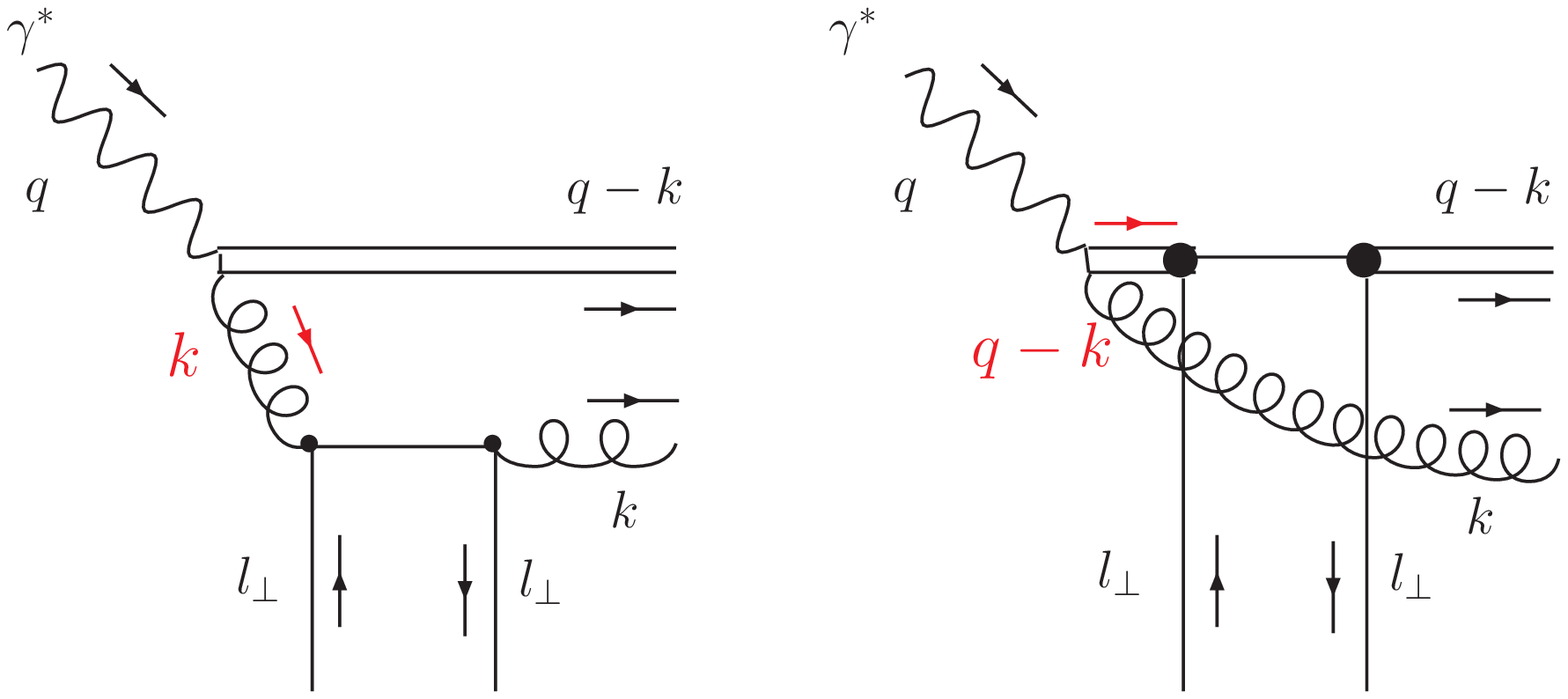}
  \end{minipage}
  \caption{The two different permutations of the couplings of the two $t$-channel sea quarks to the two components of (a) the quark dipole and (b) the effective gluon dipole.}
  \label{fig:permS}
\end{figure}
The quark dipole and effective gluon dipole interacting with this two-quark Pomeron are shown in Fig.~\ref{fig:dipoleS}.  The light-cone wave functions of the photon, $\Psi(\alpha,\vec{k_t})$, are the same as those given in Section \ref{sec:two-gluon-exchange}.  The two different permutations of the couplings of the two sea quarks to the two components of the dipole, shown in Fig.~\ref{fig:permS}, are obtained with
\begin{equation}
  D\Psi(\alpha,\vec{k_t}) \equiv 2\Psi(\alpha,\vec{k_t}),
\end{equation}
that is, there are no terms with a shifted argument, unlike for the two-gluon Pomeron.  Since there is no $\vec{l_t}$ dependence here, the integrals over $\vec{l_t}$ in the dipole factorisation formulae can be done immediately.
\begin{figure}
  \centering
  \begin{minipage}{0.5\textwidth}
    \hspace{0.35\textwidth}(a)
    \begin{center}
      \includegraphics[width=0.8\textwidth]{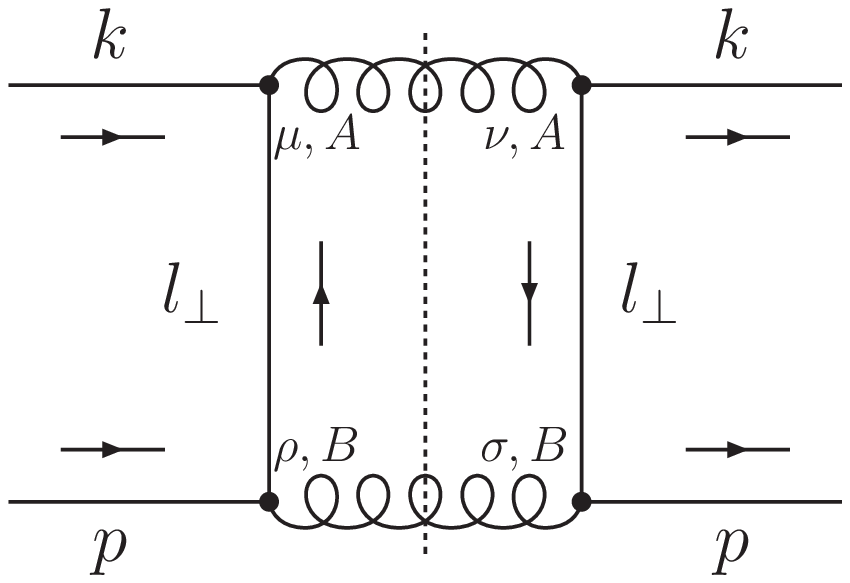}
    \end{center}
  \end{minipage}%
  \begin{minipage}{0.5\textwidth}
    \hspace{0.35\textwidth}(b)
    \begin{center}
      \includegraphics[width=0.8\textwidth]{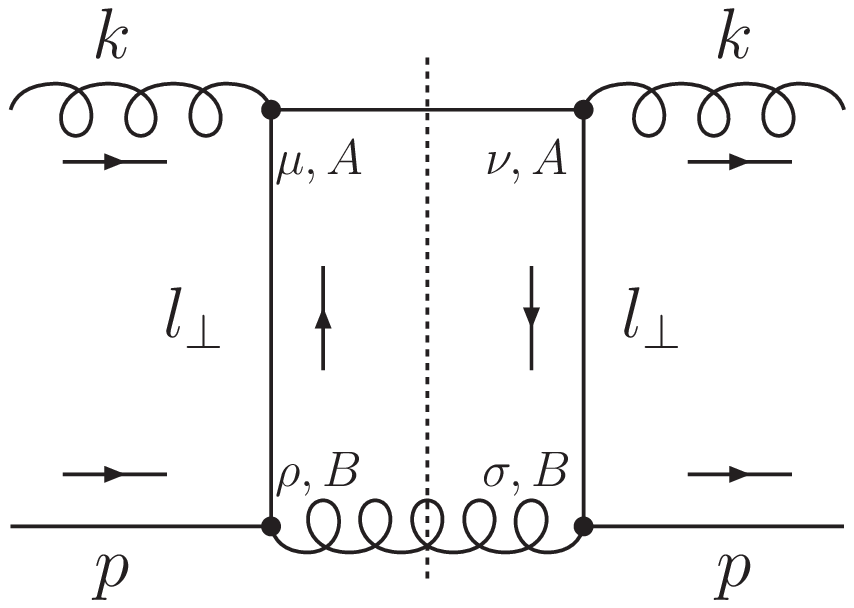}
    \end{center}
  \end{minipage}
  \caption{Cut diagrams giving the dipole cross sections for the two-quark Pomeron: (a) $qq\to gg$ and (b) $gq\to qg$.}
  \label{fig:SdipoleX}
\end{figure}

\subsubsection{Quark dipole with a transversely polarised photon}
The dipole factorisation formula for Fig.~\ref{fig:dipoleS}(a) with a transversely polarised photon is
\begin{equation} \label{eq:SqqT}
  \left.\frac{\dif\sigma_{T,q\bar{q}}^{\gamma^*p}}{\dif t}\right\rvert_{t=0} = \frac{N_C}{16\pi}\int_0^1\!\dif\alpha\int\!\frac{\dif k_t^2}{2\pi}\sum_f e_f^2\,\alpha_{\mathrm{em}}\;\frac{1}{2}\sum_{h,h^\prime,\gamma=\pm1}\left\lvert D\Psi_{hh^\prime}^\gamma\right\rvert^2\;\hat{\sigma}^2,
\end{equation}
where
\begin{equation} \label{eq:SqqTwave}
  \sum_{\gamma,h,h^\prime=\pm1} \left\lvert D\Psi_{hh^\prime}^\gamma(\alpha,\vec{k_t})\right\rvert^2 = \frac{16}{\mu^4}\left\{\left[(1-\beta)\mu^2-m_f^2\right]\left(1-2\beta\frac{\mu^2}{Q^2}\right)+m_f^2\right\}.
\end{equation}
The dipole cross section for $qp\to gp$ is obtained from the scattering process $qq\to gg$ with $t$-channel sea-quark exchange, shown in Fig.~\ref{fig:SdipoleX}(a).  Here, the squared matrix element is
\begin{equation}
  |\mathcal{M}|^2 = \frac{1}{4}\mathcal{C}\;\frac{g^4}{l_t^4}\;\mathrm{Tr}[\gamma^\mu\slashed{k}\gamma^\nu\slashed{l}_\perp\gamma^\sigma\slashed{p}\gamma^\rho\slashed{l}_\perp]\;d_{\mu\nu}(k+l_\perp,p)\;d_{\rho\sigma}(p-l_\perp,k),
\end{equation}
where the colour factor is
\begin{equation}
  \mathcal{C}(qq\to qq) = \frac{1}{N_C^2}\mathrm{Tr}[t^At^At^Bt^B] = \frac{16}{27}.
\end{equation}
In the high-energy limit, where only the terms to leading $\mathcal{O}(\hat{s}/|\hat{t}|)$ are retained,
\begin{equation}
  \frac{\dif\hat\sigma}{\dif l_t^2}(qq\to gg) = \frac{1}{l_t^2}\frac{32\pi}{27}\frac{\xPom}{\mu^2}\alpha_S(l_t^2)\alpha_S(\mu^2).
\end{equation}
Note that this expression is suppressed by an extra factor $\mathcal{O}(|\hat{t}|/\hat{s})$ compared to the $t$-channel gluon exchange processes considered in Section \ref{sec:two-gluon-exchange}, where $\hat{s}\simeq\mu^2/\xPom$ and $\hat{t}=-l_t^2$.  We obtain the cross section for $qp\to gp$ by making the replacement
\begin{equation} \label{eq:Pqq-fq}
  \left.\frac{\alpha_S(l_t^2)}{2\pi}\xPom P_{qq}(\xPom)\right\rvert_{\xPom\ll 1} = \frac{\alpha_S(l_t^2)}{2\pi}\,\xPom\,C_F=\frac{2}{3\pi}\xPom\alpha_S(l_t^2)\to f_q(\xPom,l_t^2,\mu^2),
\end{equation}
where $f_q(\xPom,l_t^2,\mu^2)$ is the unintegrated quark distribution of the proton.  This replacement gives
\begin{equation}
  \hat{\sigma}(qp\to gp) = \frac{16\pi^2}{9}\frac{\alpha_S(\mu^2)}{\mu^2}\int_0^{\mu^2}\!\diff{l_t^2}f_q(\xPom,l_t^2,\mu^2) = \frac{16\pi^2}{9}\frac{\alpha_S(\mu^2)}{\mu^2}\xPom q(\xPom,\mu^2),
\end{equation}
where $q(\xPom,\mu^2)$ is the integrated quark distribution of the proton.  Again, we should use the \emph{skewed} quark distributions, which gives rise to an extra factor \cite{Shuvaev:1999ce}
\begin{equation} \label{eq:Rq}
  R_q(\lambda) = \frac{2^{2\lambda+3}}{\sqrt{\pi}}\frac{\Gamma(\lambda+5/2)}{\Gamma(\lambda+3)},
\end{equation}
assuming that $\xPom\,q(\xPom,\mu^2)\propto \xPom^{-\lambda}$ at small $\xPom$.
The final formula for massive quarks is
\begin{equation} \label{eq:Sftd3mass}
  F_{T,q\bar{q}}^{{\rm D}(3)} = \sum_f e_f^2\int_{\frac{m_f^2}{1-\beta}}^{\frac{Q^2}{4\beta}}\!\diff{\mu^2}\;\frac{1}{\sqrt{1-4\beta\mu^2/Q^2}}\frac{1}{\xPom B_D} \left[\,R_q\frac{\alpha_S(\mu^2)}{\mu}\;\xPom q(\xPom,\mu^2)\,\right]^2\;\frac{16\beta}{27\mu^2}\left\{\ldots\right\},
\end{equation}
where $\left\{\ldots\right\}$ denotes everything inside the curly brackets in \eqref{eq:SqqTwave}.  Replacing $\sum_fe_f^2$ by $e_c^2$ and $q(\xPom,\mu^2)$ by $c(\xPom,\mu^2)$ in \eqref{eq:Sftd3mass} gives a formula for diffractive open charm production.  Alternatively, assuming light quark flavour symmetry, $q(\xPom,\mu^2) = S(\xPom,\mu^2)/(2n_f)$ with $n_f=3$, and making the approximation $\mu^2\ll Q^2$ we obtain
\begin{equation} \label{eq:f2d3SdipoleT}
  F_{T,q\bar{q}}^{{\rm D}(3)} = \int_{\mu_0^2}^{Q^2}\!\diff{\mu^2}\;f_{\Pom=S}(\xPom;\mu^2)\;F_T^{\Pom=S}(\beta,\mu^2;\mu^2),
\end{equation}
where the `Pomeron flux factor' is\footnote{Just as in \eqref{eq:f2d3GdipoleTPomflux}, we use a slightly different definition from that given in \cite{MRW1,MRW2}.  Since only the combination \eqref{eq:f2d3SdipoleT} matters, we are free to redistribute factors of $\mu$ and constants as we please.}
\begin{equation} \label{eq:Seafluxfactor}
  f_{\Pom=S}(\xPom;\mu^2) = \frac{1}{\xPom B_D} \left[\,R_S\frac{\alpha_S(\mu^2)}{\mu}\;\xPom S(\xPom,\mu^2)\,\right]^2,
\end{equation}
and the `Pomeron structure function' at a scale $\mu$ originating from a component of the Pomeron of size $1/\mu$ is
\begin{equation} \label{eq:qS}
  F_T^{\Pom=S}(\beta,\mu^2;\mu^2) = \langle e_f^2\rangle\beta\Sigma^{\Pom=S}(\beta,\mu^2;\mu^2) = \langle e_f^2\rangle\frac{4}{81}~\beta(1-\beta).
\end{equation}
Recall that the notation $\Pom=S$ is used to indicate that the perturbative Pomeron is represented by two $t$-channel sea quarks.  Again, \eqref{eq:qS} can be taken as the initial condition for DGLAP evolution and it can be regarded as the Pomeron-to-quark splitting function, $P_{q,\Pom=S}$; see \eqref{eq:qSa}.

\subsubsection{Quark dipole with a longitudinally polarised photon}
The dipole factorisation formula for Fig.~\ref{fig:dipoleS}(a) with a longitudinally polarised photon is
\begin{equation}
  \left.\frac{\dif\sigma_{L,q\bar{q}}^{\gamma^*p}}{\dif t}\right\rvert_{t=0} = \frac{N_C}{16\pi}\int_0^1\!\dif\alpha\int\!\frac{\dif k_t^2}{2\pi}\sum_f e_f^2\,\alpha_{\mathrm{em}}\;\sum_{h,h^\prime=\pm1}\left\lvert D\Psi_{hh^\prime}^{\gamma=0}\right\rvert^2\hat{\sigma}^2,
\end{equation}
where
\begin{equation}
  \sum_{h,h^\prime=\pm1} \left\lvert D\Psi_{hh^\prime}^{\gamma=0}(\alpha,\vec{k_t})\right\rvert^2 = \frac{32}{\mu^2}\frac{\mu^2}{Q^2}\beta^2,
\end{equation}
leading, in the case of massive quarks, to
\begin{equation} \label{eq:Sfld3mass}
  F_{L,q\bar{q}}^{{\rm D}(3)} = \sum_f e_f^2\int_{\frac{m_f^2}{1-\beta}}^{\frac{Q^2}{4\beta}}\!\diff{\mu^2}\;\frac{1}{\sqrt{1-4\beta\mu^2/Q^2}}\frac{\mu^2}{Q^2}\frac{1}{\xPom B_D} \left[\,R_q\frac{\alpha_S(\mu^2)}{\mu}\;\xPom q(\xPom,\mu^2)\,\right]^2\;\frac{64}{27}\beta^3,
\end{equation}
and in the limit of massless quarks, in the approximation $\mu^2\ll Q^2$, to
\begin{equation} \label{eq:LS}
  F_{L,q\bar{q}}^{{\rm D}(3)} = \frac{Q^2}{4\pi^2\alpha_{\mathrm{em}}}\frac{\beta}{\xPom}\frac{\dif\sigma_{L,q\bar{q}}^{\gamma^*p}}{\dif\beta} = \left(\int_{\mu_0^2}^{Q^2}\!\diff{\mu^2}\;\frac{\mu^2}{Q^2}\;f_{\Pom=S}(\xPom;\mu^2)\right)\;F_L^{\Pom=S}(\beta),
\end{equation}
where $F_L^{\Pom=S}(\beta)=(16/81)\,\langle e^2_f\rangle\,\beta^3$.  Note that this contribution to $\fd$ is twist-four due to the extra factor $\mu^2/Q^2$ with respect to \eqref{eq:f2d3SdipoleT}, and hence we do not perform leading-twist DGLAP evolution.

\subsubsection{Gluon dipole with a transversely polarised photon}
The dipole factorisation formula for Fig.~\ref{fig:dipoleS}(b) with a transversely polarised photon is
\begin{equation}
  \left.\frac{\dif\sigma_{gg,T}^{\gamma^*p}}{\dif t}\right\rvert_{t=0} = \frac{N_C^2-1}{16\pi}\int_0^1\!\dif\alpha\int\!\frac{\dif k_t^2}{2\pi}\sum_f e_f^2\,\alpha_{\mathrm{em}}\sum_{m,n=1,2}\left\lvert D\Psi^{mn}\right\rvert^2\hat{\sigma}^2,
\end{equation}
where
\begin{equation}
  \sum_{m,n=1,2}\left\lvert D\Psi^{mn}(\alpha,\vec{k_t})\right\rvert^2 = \frac{8}{\mu^2}(1-\beta)^2\frac{1}{\beta}.
\end{equation}
The squared matrix element for $gq\to qg$, shown in Fig.~\ref{fig:SdipoleX}(b), is
\begin{equation}
  |\mathcal{M}|^2 = \frac{1}{4}\mathcal{C}\;\frac{g^4}{l_t^4}\;\mathrm{Tr}[\gamma^\mu\slashed{l}_\perp\gamma^\rho\slashed{p}\gamma^\sigma\slashed{l}_\perp\gamma^\nu(\slashed{k}+\slashed{l}_\perp)]\;d_{\mu\nu}(k,p)\;d_{\rho\sigma}(p-l_\perp,k),
\end{equation}
where the colour factor is
\begin{equation}
  \mathcal{C}(gq\to gq) = \frac{1}{N_C}\frac{1}{N_C^2-1}\mathrm{Tr}[t^At^At^Bt^B] = \frac{2}{9}.
\end{equation}
The dipole cross section is then
\begin{equation}
  \frac{\dif\hat\sigma}{\dif l_t^2}(gq\to qg) = \frac{1}{l_t^2}\frac{32\pi}{27}\frac{\xPom}{\mu^2}\alpha_S(l_t^2)\alpha_S(\mu^2),
\end{equation}
so, making the replacement \eqref{eq:Pqq-fq}, the $gp\to qp$ cross section is
\begin{equation}
  \hat{\sigma}(gp\to qp) = \frac{2\pi^2}{3}\frac{\alpha_S(\mu^2)}{\mu^2}\int_0^{\mu^2}\!\diff{l_t^2}\sum_q f_q(\xPom,l_t^2,\mu^2) = \frac{2\pi^2}{3}\frac{\alpha_S(\mu^2)}{\mu^2}\xPom S(\xPom,\mu^2),
\end{equation}
assuming light quark flavour symmetry, $q(\xPom,\mu^2) = S(\xPom,\mu^2)/(2n_f)$ with $n_f=3$.  Again, we account for the $g\to q\bar{q}$ splitting using \eqref{eq:gtoqqbar}.  Putting everything together, we finally obtain
\begin{equation}
  F_{T,(q\bar{q})g}^{{\rm D}(3)} = \int_{\mu_0^2}^{Q^2}\!\diff{\mu^2}\;f_{\Pom=S}(\xPom;\mu^2)\;F_T^{\Pom=S}(\beta,Q^2;\mu^2),
\end{equation}
where
\begin{align}
  F_T^{\Pom=S}(\beta,Q^2;\mu^2) &= \langle e_f^2\rangle\beta\Sigma^{\Pom=S}(\beta,Q^2;\mu^2) \notag\\
  &= 2\sum_fe_f^2\frac{\alpha_S(Q^2)}{2\pi}\ln\left(\frac{Q^2}{\mu^2}\right)\beta\int_\beta^1\!\frac{\dif{\beta^\prime}}{{\beta^\prime}^2}P_{qg}\left(\frac{\beta}{\beta^\prime}\right)\,\beta^\prime g^{\Pom=S}(\beta^\prime,\mu^2;\mu^2),
\end{align}
with
\begin{equation} \label{eq:gS}
  \beta^\prime g^{\Pom=S}(\beta^\prime,\mu^2;\mu^2) = \frac{1}{9}\;(1-\beta^\prime)^2.
\end{equation}
We take \eqref{eq:gS} at a scale $\mu^2$, which can be interpreted as the Pomeron-to-gluon splitting function $P_{g,\Pom=S}$ \eqref{eq:gSa}, as the initial condition for DGLAP evolution up to $Q^2$.

\subsection{Interference between two-gluon and two-quark exchange ($\Pom=GS$)} \label{sec:interference}

We must account for interference between two-gluon and two-quark exchange, that is, interference between Fig.~\ref{fig:dipoleG}(a) and Fig.~\ref{fig:dipoleS}(a), and between Fig.~\ref{fig:dipoleG}(b) and Fig.~\ref{fig:dipoleS}(b).  We label these contributions using the notation $\Pom=GS$.

\subsubsection{Quark dipole with a transversely polarised photon}

The dipole factorisation formula for interference between Fig.~\ref{fig:dipoleG}(a) and Fig.~\ref{fig:dipoleS}(a), with a transversely polarised photon, is
\begin{equation}
  \left.\frac{\dif\sigma_{T,q\bar{q}}^{\gamma^*p}}{\dif t}\right\rvert_{t=0} = \frac{N_C}{16\pi}\int_0^1\!\dif\alpha\int\!\frac{\dif k_t^2}{2\pi}\sum_f e_f^2\,\alpha_{\mathrm{em}}\;\frac{1}{2}\sum_{\gamma,h,h^\prime=\pm1}
2\left.\left(\int\!\frac{\dif^2\vec{l_t}}{\pi}\;D\Psi_{hh^\prime}^\gamma\;\frac{\dif\hat{\sigma}}{\dif l_t^2}\right)\right\rvert_{\Pom=G}
\left.\left(D\Psi_{hh^\prime}^\gamma\;\hat{\sigma}\right)\right\rvert_{\Pom=S}.\label{eq:GSqqT}
\end{equation}
Proceeding as before, the final result for massive quarks is
\begin{multline}
    F_{T,q\bar{q}}^{{\rm D}(3)} = \sum_fe_f^2\int_{\frac{m_f^2}{1-\beta}}^{\frac{Q^2}{4\beta}}\!\diff{\mu^2}\;\frac{1}{\sqrt{1-4\beta\mu^2/Q^2}}\frac{R_gR_q}{\xPom B_D} \left[\frac{\alpha_S(\mu^2)}{\mu}\right]^2\;2\,\xPom g(\xPom,\mu^2)\,\xPom q(\xPom,\mu^2)\\\times\frac{4\beta}{9\mu^4}\left\{\left(m_f^2+\beta\mu^2\right)\left[(1-\beta)\mu^2-m_f^2\right]\left(1-2\beta\frac{\mu^2}{Q^2}\right)+m_f^2\left[m_f^2+\left(\beta-\frac{1}{2}\right)\mu^2\right]\right\}.
\end{multline}
In the limit of massless quarks, in the approximation $\mu^2\ll Q^2$, this expression reduces to
\begin{equation} \label{eq:f2d3GSdipoleT}
  F_{T,q\bar{q}}^{{\rm D}(3)} = \int_{\mu_0^2}^{Q^2}\!\diff{\mu^2}\;f_{\Pom=GS}(\xPom;\mu^2)\;F_T^{\Pom=GS}(\beta,\mu^2;\mu^2),
\end{equation}
where the `Pomeron flux factor' is
\begin{equation}
  f_{\Pom=GS}(\xPom;\mu^2) = \frac{R_gR_S}{\xPom B_D} \left[\frac{\alpha_S(\mu^2)}{\mu}\right]^2\;2\,\xPom g(\xPom,\mu^2)\,\xPom S(\xPom,\mu^2),
\end{equation}
and the `Pomeron structure function' at a scale $\mu$ originating from a component of the Pomeron of size $1/\mu$ is
\begin{equation} \label{eq:qGS}
  F_T^{\Pom=GS}(\beta,\mu^2;\mu^2) = \langle e_f^2\rangle\beta\Sigma^{\Pom=GS}(\beta,\mu^2;\mu^2) = \langle e_f^2\rangle\frac{2}{9}~\beta^2(1-\beta).
\end{equation}
Again, \eqref{eq:qGS} can be taken as the initial condition for DGLAP evolution and it can be regarded as the Pomeron-to-quark splitting function, $P_{q,\Pom=GS}$; see \eqref{eq:qGSa}.

\subsubsection{Quark dipole with a longitudinally polarised photon}

The dipole factorisation formula for interference between Fig.~\ref{fig:dipoleG}(a) and Fig.~\ref{fig:dipoleS}(a), with a longitudinally polarised photon, is
\begin{equation}
  \left.\frac{\dif\sigma_{L,q\bar{q}}^{\gamma^*p}}{\dif t}\right\rvert_{t=0} = \frac{N_C}{16\pi}\int_0^1\!\dif\alpha\int\!\frac{\dif k_t^2}{2\pi}\sum_f e_f^2\,\alpha_{\mathrm{em}}\;\sum_{h,h^\prime=\pm1}
2\left.\left(\int\!\frac{\dif^2\vec{l_t}}{\pi}\;D\Psi_{hh^\prime}^{\gamma=0}\;\frac{\dif\hat{\sigma}}{\dif l_t^2}\right)\right\rvert_{\Pom=G}
\left.\left(D\Psi_{hh^\prime}^{\gamma=0}\;\hat{\sigma}\right)\right\rvert_{\Pom=S}.\label{eq:GSqqL}
\end{equation}
For the case of massive quarks, this leads to
\begin{multline}
  F_{L,q\bar{q}}^{{\rm D}(3)} = \sum_fe_f^2\int_{\frac{m_f^2}{1-\beta}}^{\frac{Q^2}{4\beta}}\!\diff{\mu^2}\;\frac{1}{\sqrt{1-4\beta\mu^2/Q^2}}\frac{\mu^2}{Q^2}\frac{R_gR_q}{\xPom B_D} \left[\frac{\alpha_S(\mu^2)}{\mu}\right]^2\;2\,\xPom g(\xPom,\mu^2)\,\xPom q(\xPom,\mu^2)\\\times\frac{8\beta^3}{9}\left(2\beta-1+\frac{2m_f^2}{\mu^2}\right),
\end{multline}
and in the limit of massless quarks, in the approximation $\mu^2\ll Q^2$, we obtain
\begin{equation}
  F_{L,q\bar{q}}^{{\rm D}(3)} = \frac{Q^2}{4\pi^2\alpha_{\mathrm{em}}}\frac{\beta}{\xPom}\frac{\dif\sigma_{L,q\bar{q}}^{\gamma^*p}}{\dif\beta} = \left(\int_{\mu_0^2}^{Q^2}\!\diff{\mu^2}\;\frac{\mu^2}{Q^2}\;f_{\Pom=GS}(\xPom;\mu^2)\right)\;F_L^{\Pom=GS}(\beta),
\end{equation}
where $F_L^{\Pom=GS}(\beta)=(4/9)\,\langle e^2_f\rangle\,\beta^3\left(2\beta-1\right)$.  Note that this contribution to $\fd$ is twist-four due to the extra factor $\mu^2/Q^2$ with respect to \eqref{eq:f2d3GSdipoleT}, and hence we do not perform leading-twist DGLAP evolution.

\subsubsection{Gluon dipole with a transversely polarised photon}

The dipole factorisation formula for interference between Fig.~\ref{fig:dipoleG}(b) and Fig.~\ref{fig:dipoleS}(b), with a transversely polarised photon, is
\begin{equation}
  \left.\frac{\dif\sigma_{gg,T}^{\gamma^*p}}{\dif t}\right\rvert_{t=0} = \frac{N_C^2-1}{16\pi}\int_0^1\!\dif\alpha\int\!\frac{\dif k_t^2}{2\pi}\sum_f e_f^2\,\alpha_{\mathrm{em}}\sum_{m,n=1,2}
  2\left.\left(\int\!\frac{\dif^2\vec{l_t}}{\pi}\;D\Psi^{mn}\;\frac{\dif\hat{\sigma}}{\dif l_t^2}\right)\right\rvert_{\Pom=G}
  \left.\left(D\Psi^{mn}\;\hat{\sigma}\right)\right\rvert_{\Pom=S}.
\end{equation}
Again, we account for the $g\to q\bar{q}$ splitting using \eqref{eq:gtoqqbar}.  Putting everything together, we finally obtain
\begin{equation}
  F_{T,(q\bar{q})g}^{{\rm D}(3)} = \int_{\mu_0^2}^{Q^2}\!\diff{\mu^2}\;f_{\Pom=GS}(\xPom;\mu^2)\;F_T^{\Pom=GS}(\beta,Q^2;\mu^2),
\end{equation}
where
\begin{align}
  F_T^{\Pom=GS}(\beta,Q^2;\mu^2) &= \langle e_f^2\rangle\beta\Sigma^{\Pom=GS}(\beta,Q^2;\mu^2) \notag\\
  &= 2\sum_fe_f^2\frac{\alpha_S(Q^2)}{2\pi}\ln\left(\frac{Q^2}{\mu^2}\right)\beta\int_\beta^1\!\frac{\dif{\beta^\prime}}{{\beta^\prime}^2}P_{qg}\left(\frac{\beta}{\beta^\prime}\right)\,\beta^\prime g^{\Pom=GS}(\beta^\prime,\mu^2;\mu^2),
\end{align}
with
\begin{equation} \label{eq:gGS}
  \beta^\prime g^{\Pom=GS}(\beta^\prime,\mu^2;\mu^2) = \frac{1}{4}\;(1-\beta^\prime)^2(1+2\beta^\prime).
\end{equation}
We take \eqref{eq:gGS} at a scale $\mu^2$, which can be interpreted as the Pomeron-to-gluon splitting function $P_{g,\Pom=GS}$ \eqref{eq:gGSa}, as the initial condition for DGLAP evolution up to $Q^2$.

\end{document}